\documentclass[aps,prb,twocolumn,superscriptaddress,amsmath,amssymb,floatfix,longbibliography]{revtex4-1}
\usepackage{bm}
\usepackage{xcolor}
\usepackage{graphicx}
\usepackage{lineno}
\usepackage[caption=false]{subfig} 
\def\ai{\textit{ab initio}}
\def\hbn{\textit{h}-BN}
\def\ks{Kohn-Sham}
\def\efield{\boldsymbol{\mathcal{E}}} 
\newcommand{\PP}{\textbf{P}}
\newcommand{\bra}[1]{\langle #1|}
\newcommand{\ket}[1]{|#1\rangle}
\newcommand{\braket}[2]{\langle #1|#2\rangle}
\begin{document}

\title{Two-photon absorption in two-dimensional materials: \\The case of hexagonal boron nitride}

\newcommand{\cinam}{CNRS/Aix-Marseille Universit\'e, Centre Interdisciplinaire de Nanoscience de Marseille UMR 7325 Campus de Luminy, 13288 Marseille Cedex 9, France}
\newcommand{\qub}{School of Mathematics and Physics, Queen's University Belfast, Belfast BT7 1NN, Northern Ireland, UK}
\newcommand{\etsf}{European Theoretical Spectroscopy Facilities (ETSF)}
\newcommand{\lem}{LEM, UMR 104 CNRS-ONERA, Universit\'e Paris Saclay F-92322 Ch{\^a}tillon, France}
\newcommand{\torvergata}{University of Rome Tor Vergata, Rome, Italy}

\author{Claudio Attaccalite}
\affiliation{\cinam}
\affiliation{\torvergata}
\author{Myrta Gr\"uning}
\affiliation{\qub}
\affiliation{\etsf}
\author{Hakim Amara}
\affiliation{\lem}
\author{Sylvain Latil}
\affiliation{SPEC, CEA, CNRS, Universit\'e Paris-Saclay, CEA Saclay 91191 Gif sur Yvette, France}
\author{Fran\c cois Ducastelle}
\affiliation{\lem}

\begin{abstract}
  We calculate the two-photon absorption in bulk and single layer hexagonal boron nitride (\hbn) both by an \ai\ real-time Bethe-Salpeter approach and by a the real-space solution of the excitonic problem in tight-binding formalism.
The two-photon absorption obeys different selection rules from those governing linear optics and therefore provides complementary information on the electronic excitations of \hbn. Combining the results from the simulations with a symmetry analysis we show that two-photon absorption is able to probe the lowest energy $1s$ states in the single layer \hbn\ and the lowest dark degenerate dark states of bulk \hbn . This deviation from the ``usual'' selection rules based on the continuous hydrogenic model is explained within a simple model that accounts for the crystalline symmetry. The same model can be applied to other two-dimensional materials with the same point-group symmetry, such as the transition metal chalcogenides. We also discuss the selection rules related to the inversion symmetry of the bulk layer stacking.
\end{abstract}           

\maketitle

\section{Introduction}
Two-photon absorption (TPA) is a nonlinear optical process in which the absorption of two photons excites a system to a higher energy electronic state. Two-photon transitions obey selection rules distinct from those governing linear excitation processes and thereby provide complementary insights into the electronic structure of excited states, as has been demonstrated in molecular systems\cite{Callis1997} and bulk solids.\cite{Shen1984}  
In particular it is frequently argued that one-photon processes are only allowed for excitons of $s$ symmetry whereas $p$ states can be observed in TPA. These rules can be derived within a continuous hydrogenic  model for excitons where full rotational symmetry is assumed. They have been invoked to analyze the excitonic effects observed in carbon nanotubes,\cite{Wang2005} and also recently in bulk \hbn.\cite{Cassabois2016} Actually these rules are not generally valid if the genuine crystalline symmetry is taken into account, but they can at least provide interpretations in terms of high or low oscillator strength.\cite{Barros2006}

Nonlinear optical properties of two-dimensional (2D) crystals, and as such the TPA, have been recently  the object of several experimental and computational studies. For example, a giant TPA has been reported\cite{li2015giant,zhang2015direct} for transition metal dichalcogenides (TMDs)  which has been attributed to the peculiar optical properties of 2D crystals; in another study on TMDs, TPA has been used to probe excited states which are dark in linear optics.\cite{he2014tightly}

As a consequence, there is a revival of interest about the selections rules for two-photon transitions.\cite{Gang2017} In TMDs, the symmetry of the excitonic states is related to the existence of gaps at the $K$ points of the Brillouin zone.\cite{Kang2016} Excitonic effects are strong and the exciton wave functions are fairly localized, so that the low threefold symmetry plays an important role.\cite{Wu2015} Although it has been first argued that the usual selection rules based on the hydrogenic model are also valid,\cite{Berkelbach2015} more accurate studies have shown that this is actually not the case, for one-photon as well as for two-photon processes.\cite{Xiao2015,Galvani2016,Glazov2017,Gong2017}

Here we analyze the case of the  \hbn\ single layer and bulk, which have  the same lattice symmetry as the TMDs and very strongly bound excitons. We combine tight-binding calculations\cite{Galvani2016} of the two-photon transition probability with sophisticated \ai\ real-time Bethe-Salpeter simulations\cite{Attaccalite2013} of the two-photon resonance third-order susceptibility. This combination is a unique feature of this work: on the one hand the tight-binding calculations allow us to identify the symmetry properties of the excitons, on the other hand the \ai\ real-time Bethe-Salpeter simulations provide the TPA spectra---to our knowledge the first \ai\ TPA spectra at this level of theory---which can be compared quantitatively with experiment. These two very different approaches show consistently that the TPA is able to probe the lowest $1s$ exciton in the bulk and single layer \hbn .

In Sec.~\ref{sec:choice} we discuss the choice of \hbn\ as a case study and describe the tight-binding modeling of its electronic and optical properties. In Sec.~\ref{sec:tpa} we detail how the two-photon transition probabilities and the  two-photon resonance third-order susceptibility are obtained respectively within the tight-binding and the \ai\ framework. We then show and compare the results of the \ai\ real-time simulations (Sec.~\ref{sec:resab}) and of the tight-binding calculations (Sec.~\ref{sec:restb}) for the single layer \hbn. We also contrast the case of the single layer with the bulk, highlighting the role of the inversion symmetry. Finally, we discuss  the selection rules for one- and two-photon processes and on the basis of our results we clarify few recent experimental works on nonlinear optical properties of 2D crystals and bulk \hbn (Sec.~\ref{sec:rules}).


\section{Electronic, optical properties and tight binding model of {\lowercase{\textit{h}}-BN }}
\label{sec:choice}
\subsection{Choice of the system}
We have chosen \hbn\ monolayer as a case study for several reasons. One reason is the abundance of experimental studies (luminescence,\cite{Watanabe2009,Watanabe2011,Jaffrennou2007,Museur2011,Pierret2014,Schue2016,Li2016,Cassabois2016,Vuong2017} X-rays,\cite{Galombosi2011,Fugallo2015} or electron energy loss spectroscopy.\cite{Fossard2017,Schuster2017,Sponza2017}) on the electronic and optical properties of both the bulk and the single layer. 
These studies, supported by theoretical investigations,  \cite{Arnaud2006,Arnaud2008,Wirtz2006,Wirtz2008,Schue2018,Paleari2018}  have shown that \hbn\ is an indirect band gap insulator whose optical absorption spectrum  is dominated by strongly bound excitons. Then, the electronic structures of the \hbn\ monolayer and of the bulk structure are fairly well known---which is convenient for the tight-binding model.\cite{Blase1995,Arnaud2006} A \hbn\ monolayer is simply the honeycomb lattice where B and N atoms alternate on the hexagons.  The bands close to the gap are built from the $\pi$ states. In the case of the monolayer,  there are just  a single valence $\pi$ band and a conduction $\pi^*$ band which are nearly parallel along the $KM$ direction of the Brillouin zone. The gap is direct at $K$ point and about 7 eV. \cite{Galvani2016}

It is of interest as well to compare the TPA in  the single layer and in the layered bulk system. The stablest bulk structure corresponds to a so-called $AA'$ stacking where B and N atoms alternate along lines parallel to the stacking axis. The periodicity along this axis is  twice the inter-planar distance and the lattice parameters used in our simulation are $a= 2.5$~\AA\ and $c/a=2.6$.  \cite{solozhenko1995isothermal} In the bulk,  there are then two $\pi$ and two $\pi^*$ bands and the gap becomes indirect between a point close to $K$ (valence band) and $M$ (conduction band).\cite{Kang2016}

Another reason for choosing \hbn\ is the strong bound excitons in the absorption spectrum of both the monolayer and the bulk\cite{Wirtz2006,Galvani2016,Wirtz2008} since as mentioned before, we expect the effect of the low threefold symmetry to be more visible for very localized excitons. More practically, spectral features corresponding to strongly localized exciton converge easily with the numerical parameters within the \ai\ framework allowing for accurate and not too cumbersome calculations. 
Furthermore, while in the absorption spectrum of the single layer there is only a pair of degenerate excitons,\cite{Galvani2016} in the absorption spectrum of bulk \hbn\ there are two degenerate pairs of excitons: one dark pair, which is the lowest in energy, and one bright pair (Davydov splitting).\cite{Arnaud2006,Arnaud2008,Wirtz2008} Probing the lowest excitons in  bulk \hbn\ remains a challenge. We expect that since the system has an inversion symmetry, the dark states in the absorption become bright in the TPA and \textit{vice versa}, so that it may be possible to probe the  lowest excitons in the bulk \hbn\ in TPA. In Sec.\ref{sec:res} we show that this is indeed the case.  

Finally, other systems of interest, such as the TMDs, have the same lattice symmetry, so that results depending on the symmetry properties can be generalised  or can be used as a starting point when studying those systems.

\subsection{Tight-binding model}
\label{tboptics}

In tight-binding formalism, we used a simple model that describes top valence and bottom conduction bands close to the gap, and which  is stable to catch the most relevant physics of optical excitation in  \hbn.
 It turns out that the valence states close to the gap are almost completely centred on the nitrogen sites whereas the conduction states are centred on the boron sites. For the monolayer this has allowed us to derive a very simple tight-binding model characterized by two parameters: an atomic parameter $\Delta$ related to  the atomic levels, $+\Delta$ for boron, and $-\Delta$ for nitrogen, and a transfer integral between first neighbours $-t, t>0$. In this model the direct gap is exactly equal to $2\Delta$. Close to the gap the electronic structure can be simplified and the electrons in the conduction band (valence band) can be considered as moving on the B (N) sublattice with an effective  transfer integral equal to $t^2/2\Delta$. The Wannier states associated with the $\pi$ and $\pi^*$ bands are then mainly localized on N and B sublattices with small components $\pm (t/2\Delta)$ on the other ones. The best fit to \ai\ data are obtained with $t= 2.30$ eV, $\Delta=3.625$ eV.\cite{Galvani2016}
 
 Notice than when $\Delta = 0$ the model is that of graphene where the gap vanishes at $K$ points. This discrete tight-binding model has a continuous counterpart when expanding the equations close to $K$ in reciprocal space. The band structure is then described within a Dirac model for massless 2D electrons. Conversely, in the case of the \hbn\ monolayer we can use a continuous massive Dirac model. The simplest extension of the 2D TB model to bulk \hbn\ consists in introducing transfer integrals between nearest neighbours along the stacking axis.\cite{Kang2016,Paleari2018} 
\subsection{Optical matrix elements}

To first order (one-photon processes), the coupling with the electromagnetic field is  described using the hamiltonian $H_{I1}= -e\,\textbf{p}.\textbf{A}/m$, where $\textbf{A}$ is the vector potential varying as $e^{-i\omega t}$. To this order, $\textbf{p}/m = \textbf{v} =\frac{1}{i\hbar}[\textbf{r},H]$, where $\textbf{v}$ is the velocity operator and $H$ is the hamiltonian of the unperturbed system. Using the tight-binding scheme, where $|\textbf{n}\rangle$ denotes a $\pi$ state at site $\textbf{n}$, we have $\textbf{v}_{\textbf{nm}} = \langle\textbf{n}|\textbf{v}|\textbf{m}\rangle$ = $(\textbf{n}-\textbf{m}) t_{\textbf{nm}}/i\hbar$, where $t_{\textbf{nm}}$ is the transfer integral between sites $\textbf{n}$ and $\textbf{m}$. Using our simple model for the \hbn\ sheet, we keep only first neighbour integrals $-t$ and the matrix element couples valence states $|\textbf{m}_v\rangle$ to conduction states $|\textbf{n}_c\rangle$. Actually, as mentioned above, the atomic states should be replaced by Wannier states:
\begin{align}
|\textbf{m}_ {v}\rangle_w &\simeq |\textbf{m}_N\rangle +  \frac{t }{2\Delta} \sum_{\bm{\tau}}  |\textbf{m}_N + \bm{\tau}\rangle 
\label{Wannier1}\\
|\textbf{n}_ {c}\rangle_w &\simeq |\textbf{n}_B\rangle -  \frac{t }{2\Delta} \sum_{\bm{\tau}}  |\textbf{n}_B - \bm{\tau}\rangle  \; ,
\label{Wannier2}
\end{align}
where $|\textbf{m}_N\rangle$ and $|\textbf{n}_B\rangle$  denote the genuine atomic states centred on the N and B atoms, and where the three $\bm{\tau}$ vectors are the first neighbour vectors $\textbf{n}_B-\textbf{m}_N$. The corresponding Bloch functions are given by:
\begin{align}
|\textbf{k}_v\rangle &\simeq |\textbf{k}_N\rangle +  \frac{t }{2\Delta} f^*(\textbf{k})| \textbf{k}_B\rangle\\
|\textbf{k}_c\rangle & \simeq |\textbf{k}_B\rangle - \frac{t }{2\Delta} f(\textbf{k})| \textbf{k}_N\rangle  \; ,
\end{align}
where the $| \textbf{k}_{N(B)}\rangle$ are the Bloch functions built from the atomic orbitals, $\ket{\textbf{k\,}_{N(B)}} = \frac{1}{\sqrt{N}}\sum_{n\in N(B)} e^{i\textbf{k}.\textbf{n}}\,\ket{\textbf{n}}$, and where $f(\textbf{k}) = \sum_{\bm{\tau}} e^{i\textbf{k}.\bm{\tau}}$, and $N$ is the number of lattice cells. To  lowest order in $t/\Delta$, we have therefore:
\begin{equation}
\bra{\textbf{k}_c}\textbf{v}\,\ket{\textbf{k}_v} \simeq  \bra{\textbf{k}_B}\textbf{v}\,\ket{\textbf{k}_N} 
=-\frac{t}{\hbar} \,\nabla_{\textbf{k}} f({\textbf{k}})^*  \; .
\end{equation}
At low energy we can expand $\textbf{k}$ close to the $K$ and $K'$ (equivalent to $-K$) points. Up to a phase factor which depends on the choice of the $K$ points, we obtain:
\begin{equation}
\bra{\textbf{k}_c}\textbf{v}.\textbf{e}\,\ket{\textbf{k}_v}  \simeq \bra{\pm\textbf{K}_c}\textbf{v}.\textbf{e}\,\ket{\pm\textbf{K}_v}  \simeq 
v_F (e_x\pm ie_y) \; ,
\label{matrix_element}
\end{equation}
where $v_F=\frac{3}{2}a_{bn}t$ is the slope of $|(f(\textbf{k}+\textbf{q})|$ for small $\textbf{q}$, $|(f(\textbf{k}+\textbf{q})| \simeq v_F q$ when $q \to 0$, and $a_{bn}$ is the distance between first neighbor $B$ and $N$ atoms. $\textbf{e}$ is the vector characterizing the light polarization, where $\pm\textbf{K}_c$ are the wave functions at the $\pm K$ points.  The matrix elements are therefore maximum for circular polarizations of opposite signs in  $K$ and $K'$ valleys. This result is now fairly well known in the case of transition metal dichalcogenides but it is worth insisting  on a particular point. The optical coupling between valence and conduction bands is not related here to the symmetry of the $\pi$ states of nitrogen and boron. Indeed, when considering polarization vectors within the sheet plane, they behave as $s$ states.  The coupling is due to the fact that the Wannier functions of the $\pi$ and $\pi^*$ bands are centred on different sublattices. Since furthermore there is no symmetry centre connecting these sublattices a finite dipolar moment is allowed, of the circular type here, because of the threefold symmetry.
\footnote{In the case of dichalcogenides the symmetry analysis is more complex because of the presence of $d$ orbitals of different symmetries.}

\subsection{One-photon absorption for independent single particles }

Within a one-particle model and neglecting the photon wave vector, the one-photon absorption is proportional to the transition probability $W$ given by the Fermi golden rule:
\begin{equation*}
W = \frac{2\pi}{\hbar} \frac{e^2}{\omega^2}({\bm{\mathcal{E}}}/2)^2\sum_{\textbf{k}} |\bra{\textbf{k}c}\textbf{v}.\textbf{e}\,\ket{\textbf{k}v}|^2 \delta (E_{\textbf{k}c} - E_{\textbf{k}v} - \hbar\omega) \; ,
\end{equation*}
where ${\bm{\mathcal{E}}}$ is the electrical field amplitude. In our model the valence and conduction bands are symmetric, $E_{\textbf{k}_v} =-E_{\textbf{k}_c}$, so that $E_{\textbf{k}c}-E_{\textbf{k}v} \simeq 2\Delta +  \frac{t^2}{\Delta} |f(\textbf{k})|^2$, where $|f(\textbf{k})|^2|$ is the band energy of a triangular lattice with  transfer integrals equal to $t^2/\Delta$ and centred on $2\Delta+3t^2/\Delta$.\cite{Galvani2016} Neglecting the dependence on $\textbf{k}$ of the matrix element, and replacing $E=\hbar\omega$ by $2\Delta$ close to the gap, we recover the usual formula for the absorption, proportional to the  joint density of states, which is here also proportional to the densities of states of the valence and conduction bands.\cite{Galvani2016}

\subsection{Excitons }
In the presence of electron-hole interactions,  excitonic effects come into play. They are usually treated using a Bethe Salpeter formalism, which, within our TB formalism,  can be reduced to an effective Wannier-Schr\"{o}dinger equation for electron-hole pairs. 

\subsubsection{Formalism}

Let then $\ket{\Phi} $ be such an excitonic state. In our model for the monolayer, it can be written as:
\begin{equation}
\ket{\Phi} = \sum_{\textbf{k}}\Phi_{\textbf{k}} \; a_{\textbf{k}_c}^+ a_{\textbf{k}_v}^{} \ket{\emptyset}  \; ,
\end{equation}
where $a_{\textbf{k}_c}^+$ and $ a_{\textbf{k}_v}$ are the usual creation and annihilation operators for conduction and valence electrons and $ \ket{\emptyset}$ is the unperturbed ground state. It is  convenient to work in real space, using the Wannier basis for electrons and holes, and to write:
 \begin{equation*}
\ket{\Phi} = \sum_{\textbf{n}\textbf{m}}\Phi_{\textbf{n}\textbf{m}} \; a_{\textbf{n}_c}^+ a_{\textbf{m}_v}^{} \ket{\emptyset}  \; .
\end{equation*}
Within this formalism the action of the velocity operator on the ground state can be written:
 \begin{equation}
\textbf{v}\ket{\emptyset} =\sum_{\textbf{n},\textbf{m}} \textbf{v} _{\textbf{n}\textbf{m}}  a_{\textbf{n}_c}^+ a_{\textbf{m}_v}^{} \ket{\emptyset} 
= -\sum_{\textbf{m},\bm{\tau}} \frac{t}{i\hbar}\, \bm{\tau} \,  a^{+}_{(\textbf{m}_v+\bm{\tau})_c}   a_{\textbf{m}_v}^{} \ket{\emptyset} \; .
\label{velocity}
\end{equation}
 Finally we define electron-hole states in reciprocal and real spaces:
\begin{align}
\ket{\textbf{k}vc} &= a^+_{\textbf{k}_c} a_{\textbf{k}_v} ^{}\ket{\emptyset} = \frac{1}{\sqrt{N}}\sum_{\textbf{R}} e^{i\textbf{k}.\textbf{R}} \ket{\textbf{R}{vc}}\\
\ket{\textbf{R}{vc}} &= \frac{1}{\sqrt{N}} \sum_{\textbf{n}} a^+_{\textbf{n}_v+\textbf{R}{vc}} \; a_{\textbf{n} v}^{}  \ket{\emptyset}  \; .
\end{align}
One should be careful in distinguishing the variables associated with the motion of the electron-hole pair as a whole from those describing the relative motion of the electron and of the hole. The latter ones are  $\textbf{k}$ and  $\textbf{R}$. The position of the pair is defined here by the position of the hole, whereas  its motion as a whole is characterized by $\textbf{Q} =\textbf{k}_c-\textbf{k}_v$ which is a good quantum (conserved) number, taken here equal to 0. Actually $\ket{\textbf{R}{vc}}$ is nothing but the Bloch function for the motion of the corresponding pair for $\textbf{Q}=0$.  The extension of the formalism to $\textbf{Q}\neq 0$ will be treated elsewhere. In the following we  drop the subscripts $v$ and $c$. Eq.(\ref{velocity}) can then be rewritten as:
\begin{equation}
\textbf{v}\ket{\emptyset} = (it/\hbar) \sum_{\bm{\tau}}\bm{\tau}  \ket{\bm{\tau}} \; .
\end{equation}

 \subsubsection{Bethe-Salpeter Wannier equation}
We recall here the formalism derived in Ref.[\onlinecite{Galvani2016}]. We start from the usual simplified Bethe Salpeter equation which is just an effective Schr\"{o}dinger equation for electron-hole pairs:
\begin{equation*}
(E_{\textbf{k}_c} - E_{\textbf{k}_v})\Phi_{\textbf{k}{vc}} + \sum_{\textbf{k}'{v'c'}} \bra{\textbf{k}{vc}} K_{eh} \ket{\textbf{k}'{v'c'}} \Phi_{\textbf{k}'{v'c'}} 
= E\  \Phi_{\textbf{k}{vc}}  \; .
\end{equation*}
In this equation, $E_{\textbf{k}_c} - E_{\textbf{k}_v}$ plays the part of a band (or kinetic) energy, approximated here by  $2\Delta + t^2  |f({\textbf{k}})|^2/\Delta  $. The interaction term $K_{eh}$ contains a direct Coulomb contribution and an exchange term which will be neglected here. Coming back in real space, we obtain an excitonic hamiltonian $H_{eh}$:
\begin{align}
H_{eh} &= H^0_{eh} + U  \label{Heh}\\
\bra{\textbf{R}}H^0_{eh}\ket{\textbf{R}'} &= 2\Delta + 3t^2/\Delta\quad \text{if}\quad  \textbf{R} =\textbf{R}' \\
&= t^2/\Delta \quad \text{if }\, \textbf{R} \;\text{and} \; \textbf{R}' \; \text{are nearest neighbours}\\
U &= \sum_{\textbf{R}} \ket{\textbf{R}} U_{\textbf{R}}\bra{\textbf{R}} \; .
\end{align}
According to our definitions the vectors $\textbf{R}$ have their origin on a hole (nitrogen) site and their extremities on the electron (boron) triangular sublattice. We then see that the above hamiltonian is similar to a tight-binding hamiltonian for electrons moving on a triangular lattice in the presence of an ``impurity'' at the origin (at the centre of a triangle). If the potential $U_{\textbf{R}}$ is known, this is a classical problem which can be solved with standard techniques, using diagonalization or Green function techniques.

 \subsubsection{Optical matrix element}
 \label{opt}

In the many body excitonic language, the relevant matrix element is:
\footnote{The matrix element of the velocity operator does not depend on the Coulomb potential which has been assumed to be local, and therefore commutes with the $\textbf{r}$ operator.}
\begin{equation}
W_{\Phi} \equiv \bra{\bm{\Phi}} \textbf{e}.\textbf{v}\ket{\emptyset} =	(it/\hbar) \sum_{\tau}\textbf{e}.\bm{\tau} \braket{\bm{\Phi}}{\bm{\tau}} 	 \; .
\end{equation}
If we define a dipole matrix element $\textbf{d}_\Phi = \sum_{\tau}\bm{\tau} \braket{\bm{\tau}}{\bm{\Phi}}  $, the transition probability associated with exciton $\bm{\Phi}$ is given by:
\begin{equation}
P_{\Phi} = \frac{2\pi}{\hbar} \frac{e^2 ({\mathcal{E}}/2)^2 t^2}{E^2}  \, |\textbf{e}.\textbf{d}_\Phi	|^2	\,\delta (E-E_\Phi) \; ,
\end{equation}
where  $E_\Phi$ is the exciton energy. Since $E_\Phi \sim\Delta$, this probability is of order $(t/\Delta)^2$. In the case of the BN monolayer, the point symmetry is that of  the $C_{3v}$ group and vectors transform as the two-dimensional $E$ representation of this group. The exciton is therefore ``bright'', $\textbf{d}_\Phi	\neq 0$, if $\ket{\Phi}$ also transforms as $E$. Notice also that only the local components of the wave function are involved in $\textbf{d}_\Phi$. This is the discrete equivalent of the classical Elliott theory: Within a continuous  model the matrix element is proportional to $\Phi(\bm{r} =0)$.\cite{Yu2010,Toyozawa2003} In our case, the ground state exciton of the monolayer, discussed in detail in Ref.[\onlinecite{Galvani2016}], does transform as $E$. If the position of the hole is fixed at the origin, the exciton wave function extends principally on the first neighbours and its oscillator strength is very large. More precisely, let $\Phi_1, \Phi_2, \Phi_3$ be the three components $\braket{\bm{\tau}}{\Phi}$. It is always possible to choose a basis $(\Phi^+ , \Phi^-)$  of the $E$ representation transforming in a ``chiral'' manner when rotated by $2\pi/3$, 
$\Phi^{\pm}_\alpha \propto e^{\pm i(2\alpha\pi/3)}$. In the  language used for Wannier-Mott excitons, these two circularly polarized components can be associated with the one-dimensional representations of the $C_3$ group at $K$ and $K$' points. Hence the notation of $s$ states in each valley used in the case of TMD. Here the intervalley coupling is more important and the classification of states based on group theory is more accurate.

A more compact formulation for the probability transition $W$ is the following:
\begin{align}
W &\propto \sum_i \bra{\emptyset} \textbf{e}.\textbf{v} \ket{\Phi_i}\delta(\hbar\omega-E_i) \bra{\Phi_i}\textbf{v}.\textbf{e} \ket{\emptyset} \\
& = \textbf{e}.  \bra{\emptyset}\textbf{v} \delta(\hbar\omega-H_{eh})   \textbf{v}\ket{\emptyset}.\textbf{e}  \; ,
\end{align}
where the sum over $i$ is over all exciton states. Finally, using the Green function $G(z)=(z-H_{eh})^{-1}$, it is sufficient to calculate $\bra{\emptyset}\textbf{v}\,G(z)\,\textbf{v}\ket{\emptyset}$.\cite{Grosso2014}

\section{Two-photon absorption}
\label{sec:tpa}
The TPA is related to the nonlinearity in the attenuation of a laser beam. Considering the intensity $I$ of a beam propagating along $\hat z$, at the lowest order beyond the linear regime, the attenuation 
is a nonlinear function of $I$:
\begin{equation}
  \label{eq:tpa}
\frac{dI}{dz} = -\alpha I - \beta I^2  \; ,
\end{equation}
where $\alpha$ is the linear absorption coefficient, and $\beta$ is the two-photon absorption coefficient (see e.g. Ref.[\onlinecite{Boyd2008}])

In what follows, we detail how we evaluate the TPA from either the third-order susceptibility extracted from \ai\ real-time dynamics (Sec.~\ref{rtoptics}) and from the transition probability of a two-photon process within a second-order perturbation treatement (Sec.~\ref{2nd_order}). 

\subsection{Nonlinear susceptibilities from \ai\ real-time dynamics}
\label{rtoptics}
The TPA coefficient in Eq.~\eqref{eq:tpa} is related to the imaginary part of the third-order response function $\chi^{(3)}$\cite{Lin2007,Boyd2008}:
\begin{equation}
\beta(\omega)= \frac{3 \omega}{2 \epsilon_0 c^ 2 n_0^2(\omega) } \mathrm{Im} \left [ \chi^{(3)} (-\omega; \omega, \omega, -\omega) \right ]   \; .
\end{equation}
where $n_0(\omega)$ is the  refractive index.  In two-photon measurements, the incoming laser frequency $\omega$ is set around half of the excited state energy $\omega_0$ we want to probe $2 \omega \simeq \omega_0$.\\ 

In this frequency  region, well below the band gap, $n_0(\omega)$ is positive, monotonic and slowly varying, therefore the peaks of $\beta(\omega)$ originates only from the imaginary part of  $\chi^{(3)}$, that is the quantity we will extract from the real-time simulations.~\cite{Attaccalite2013}
In the real-time simulations, the electronic system is excited by a time-dependent monochromatic homogeneous field. The time-evolution of the system is given by the following equation of motion for the valence band states,  
\begin{eqnarray}
i\hbar  \frac{d}{dt}| v_{m\textbf{k}} \rangle &=& \left( H^{\text{MB}}_{\textbf{k}} +i \efield \cdot \tilde \partial_\textbf{k}\right) |v_{m\textbf{k}} \rangle \label{tdbse_shf}\;,
\end{eqnarray}
where $| v_{m\textbf{k}} \rangle$ is the periodic part of the Bloch states. In the r.h.s. of Eq.~\eqref{tdbse_shf}, $ H^{\text{MB}}_{\mathbf k}$ is the effective Hamiltonian derived from many-body theory that includes both the electron--hole interaction and local field effects. The specific form of $ H^{\text{MB}}_{\mathbf k}$ is presented below. The second term in Eq.~\eqref{tdbse_shf}, $\efield \cdot \tilde \partial_\textbf{k}$, describes the coupling with the external field $\efield$ in the dipole approximation. As we imposed Born-von K\'arm\'an periodic boundary conditions, the coupling takes the form of a $\textbf{k}$-derivative operator $\tilde \partial_\textbf{k}$. The tilde indicates that the operator is ``gauge covariant'' and guarantees that the solutions of Eq.~\eqref{tdbse_shf} are invariant under unitary rotations among occupied states at $\textbf{k}$ (see Ref.~[\onlinecite{Souza2004}] for more details).

We notice that we adopt here the length gauge, which presents several advantages for \ai\ simulations.\cite{Attaccalite2013} Comparing with the velocity gauge approach---that is used in the tight-binding model---the second term of Eq.~\eqref{tdbse_shf} includes both the $\textbf{A}$ and $\textbf{A}^2$ contributions present in the velocity gauge. The two gauges are equivalent as shown in Appendix~\ref{gauge_equivalence}.

From the evolution of $| v_{m\textbf{k}} \rangle$ in Eq.~\eqref{tdbse_shf} we calculate the time-dependent polarization $\PP_\parallel$  along the lattice vector $\mathbf a$ as
 \begin{equation}
 \PP_\parallel = -\frac{ef |\mathbf a| }{2 \pi \Omega_c} \text{Im log} \prod_{\textbf{k}}^{N_{\textbf{k}}-1}\ \text{det} S\left(\textbf{k} , \textbf{k} + \mathbf q\right), \label{berryP}  \; ,
 \end{equation}
where $S(\textbf{k} , \textbf{k} + \mathbf q) $ is the overlap matrix between the valence states $|v_{n\textbf{k}}\rangle$ and $|v_{m\textbf{k} + \textbf{q}}\rangle$, $\Omega_c$ is the unit cell volume,  $f$ is the spin degeneracy, $N_{\textbf{k}}$ is the number of $\textbf{k}$ points along the polarization direction, and $\mathbf q = 2\pi/(N_{\textbf{k}} {\mathbf a})$.

The polarization can be expanded in a power series of the incoming field $\efield_j$ as:
\begin{equation}
\textbf{P}_i = \chi^{(1)}_{ij} \efield_j + \chi^{(2)}_{ijk}  \efield_j \efield_k +  \chi^{(3)}_{ijkl} \efield_j \efield_k \efield_l + O(\efield^4)  \; ,
\end{equation}
where the coefficients $\chi^{(i)}$ are a function of the frequencies of the perturbing fields and of the outgoing polarization. As explained above, the two-photon absorption is proportional to the imaginary part of the two-photon resonance third-order sosceptibility $\chi^{(3)}_{ijkl} (-\omega; \omega, \omega, -\omega)$: \emph{i.e.} the outgoing polarization has the same frequency $\omega$ of the incoming laser field, as a result of the absorption of two and the emission of one virtual photons.  

 In order to extract the $\chi^{(3)}$ coefficients we resort to a technique similar to Richardson extrapolation.~\cite{Richardson1927} In practice, we perform three different simulations with the incoming electric field at frequency $\omega$  and intensities corresponding to the amplitudes $\efield$, $\efield/2$ and $\efield/4$. The polarization resulting from each simulation can be expanded in the field as:
\begin{eqnarray}
\textbf{P} \left (\efield \right)   &=& \chi^{(1)} \efield + \chi^{(2)} \efield^2 +  \chi^{(3)} \efield^3 + O(\efield^4), \\
\textbf{P} \left (\frac{\efield}{2} \right) &=& \chi^{(1)} \frac{\efield}{2} + \chi^{(2)} \frac{\efield^2}{4} +  \chi^{(3)} \frac{\efield^3}{8} + O(\efield^4), \\
\textbf{P} \left (\frac{\efield}{4} \right) &=& \chi^{(1)} \frac{\efield}{4} + \chi^{(2)} \frac{\efield^2}{16} +  \chi^{(3)} \frac{\efield^3}{64} + O(\efield^4) \; .
\end{eqnarray}
Then we combine the three polarizations so to cancel out the linear and quadratic contributions and we obtain:
\begin{equation}
\chi^{(3)} = \frac{8}{3} \frac{ \textbf{P}(\efield)   - 6 \textbf{P}(\frac{\efield}{2}) + 8 \textbf{P}(\frac{\efield}{4}) }{\efield^3}. \; .
\label{xhi3Richardson} 
\end{equation}
The calculation is repeated for all $\omega$ in the desired range of frequencies.

The level of approximation of the so-calculated susceptibilities depends of the effective Hamiltonian that appears in the r.h.s. of Eq.~\eqref{tdbse_shf}.
Here we work in the so-called time-dependent Bethe-Salpeter framework that was introduced in Ref.[\onlinecite{attaccalite2011real}]. In this framework the Hamiltonian $H^{MB}_{\mathbf k}$ reads:
\begin{equation}
  \label{eq:hmb}
  H^{MB}_{\mathbf k} \equiv H^{\text{KS}}_\textbf{k} + \Delta H_\textbf{k} + V_h(\mathbf r)[\Delta\rho] + \Sigma_{\text{SEX}}[\Delta \gamma] \;,
\end{equation}
where $H^{\text{KS}}_\textbf{k}$ is the Hamiltonian of the unperturbed (zero-field) \ks\ system~\footnote{The \ks\ system is a fictitious system of independent particles in an effective local field such that the electronic density of the physical system is reproduced, see W. Kohn and L. J. Sham, Phys. Rev. 140, A1133 (1965)}, $\Delta H_\textbf{k}$ is the scissor operator that has been applied to the \ks\ eigenvalues, the term $V_h(\mathbf r)[\Delta\rho]$ is the time-dependent Hartree potential\cite{Attaccalite2013} and is responsible for the local-field effects\cite{Adler1962} originating from system inhomogeneities. The term $\Sigma_{\text{SEX}}$ is the screened-exchange self-energy that accounts for the electron-hole interaction,\cite{Strinati1988} and is given by the convolution between the screened interaction $W$ and $\Delta \gamma$. In the same equation:
$$\Delta \rho \equiv \rho(\mathbf r;t)-\rho(\mathbf r;t=0)$$ 
is the variation of the electronic density  and:
$$\Delta \gamma \equiv \gamma(\mathbf r,\mathbf r';t) - \gamma(\mathbf r,\mathbf r';t=0)$$
is the variation of the density matrix induced by the external field $\efield$.\\

In the limit of small perturbation Eq.~\eqref{eq:hmb} and Eq.~(\ref{tdbse_shf}) reproduce the optical absorption calculated with the standard $GW$ + BSE approach,\cite{Strinati1988} as shown both analytically and numerically in Ref.~[\onlinecite{attaccalite2011real}].

\subsection{Tight binding model}
In tight-binding the second order (two-photon processes) appears in the development of  $(\textbf{p}-e\textbf{A})^2/2m$. In  second order perturbation theory with respect to $\textbf{A}$, there are two terms. The first one is related to the linear term $(\textbf{p}-e\textbf{A})$ treated in a second order perturbation theory, and the second one comes from the $A^2$ term treated to first order. The latter one is frequently considered as non relevant in the optical regime, which is not the case here as explained below. Let us begin with the first contribution.

\subsubsection{Second order perturbation theory}
\label{2nd_order}
To second order, the transition probability $P_{i\to f}$ from an initial state $\ket{i}$ towards a final state $\ket{f}$ is given by: \cite{Grynberg2010}
\begin{equation}
P_{i\to f} = \frac{2\pi}{\hbar} \left| \frac{1}{4} \sum_{j\neq i,f} \frac{W_{fj} W_{ji}}{E_i-E_j +\hbar\omega} \right |^2 \delta(E_f - E_i -2\hbar\omega) \; ,
\end{equation}
where $W_{ij}$ is the one-photon matrix element towards the intermediate state $j$. A convenient approximation consists here to replace $E_j$ by a mean energy  between the initial energy and the exciton levels close to the bottom of the conduction band.\cite{Mahan1968,Shimizu1989, Berkelbach2015} The sum in the numerator can then be freely performed, and we are back to a situation similar to the first order calculation, where now the relevant matrix element is equal to $\bra{f}W^2\ket{i}$. Since we are looking at transition close to the gap, $\hbar\omega \sim \Delta$, and the denominator is also of order $\Delta$. The relevant matrix element is  equal to $\bra{\Phi}(\textbf{v}.\textbf{e})(\textbf{v}.\textbf{e})\ket{\emptyset}$. As seen above, the first velocity operator operating on the ground state generates electron-hole states. Therefore, the second one couples different electron-hole states. Since it is related to the kinetic energy part, it operates separately on the electron state  and on the hole state, and therefore gives contributions proportional to the velocity of the one-particle states. The corresponding matrix elements in real space are equal to $-\frac{i}{\hbar}\frac{t^2}{\Delta} \textbf{a}$, where $\textbf{a}$ is a first neighbour distance on the triangular lattice. The final complete matrix element $\bra{\Phi}(\textbf{v}.\textbf{e})(\textbf{v}.\textbf{e})\ket{\emptyset}/\Delta$ is therefore of order $t^3/\Delta^2$. There is no reason that it vanishes identically for the ground state exciton. Actually the (tensorial) product of velocity operators transform as $E\times E$ which also contains $E$. In the continuous limit however, it can be shown\cite{Berkelbach2015} that   $\bra{\Phi} \bm{v}\otimes \bm{v} \ket{\emptyset}  \sim \nabla \Phi(\bm{r})|_{\bm{r}=0} $, which implies that only $p$ states are bright. We are in a typical situation where precise selection rules based on the exact crystalline symmetry allow transitions which become forbidden if an approximate (higher) spherical symmetry is assumed.\cite{Barros2006}

\subsubsection{$A^2$ term}
In principle the $A^2$ term is local and its influence is negligible in the optical regime where the wave length is much larger than interatomic distances. At least is this true when using the full hamiltonian. In band theory we project the hamiltonian on the subspace defined by the number of bands taken into account, and the correct method to include gauge-invariant coupling with the electromagnetic field is to make the so-called Peierls substitution. This generates non-local coupling at all orders in $\textbf{A}$. More precisely in our case, we make the substitution:
\begin{equation}
t_{\textbf{nm}} \to t_{\textbf{nm}} e^{i e(\bm{n}-\bm{m}).\textbf{A})/\hbar} \; .
\end{equation}
To first order, we can check that this generates the coupling $H_{I1}$ used above, \textit{i.e.} :
\begin{equation}
\bra{\textbf{n}}H_{I1}\ket{\textbf{m}} =  i \left( \frac{e}{\hbar}\right) (\bm{m}-\bm{n}).\textbf{A}\; t_{\bm{n}\bm{m}} \; ,
\end{equation}
and therefore:

\begin{align}
\bra{\emptyset} H_{I1} \ket{\Phi^\pm} &= i  \frac{eA\,t}{\hbar }  \sum_{\bm{\tau}} (\bm{\tau}.\textbf{e})\, \Phi_{\bm{\tau}}^\pm \\
\sum_{\bm{\tau}} (\bm{\tau}.\textbf{e})  \Phi_{\bm{\tau}}^\pm &= \textbf{e}.\textbf{d}_{\Phi^\pm}	
= -\frac{3}{2}a_{bn} (e_x\pm ie_y)| \Phi_{\bm{\tau}}^\pm | \; ,
\end{align}
where $a_{bn} = |\bm{\tau}|$ and $\Phi^\pm_{\bm{\tau}}$ is the amplitude $\braket{\bm{\tau}}{\Phi^\pm}$ of the circularly polarized state defined previously.
To second order, we obtain:
 \begin{equation*}
\bra{\textbf{n}}H_{I2}\ket{\textbf{m}} =  -\frac{1}{2} \left( \frac{e}{\hbar}\right)^2 [(\textbf{m}-\textbf{n}).\textbf{A}]^2 t_{\textbf{n}\textbf{m}} \; ,
\end{equation*}
from which we deduce that:
\begin{align}
\bra{\emptyset} H_{I2} \ket{\Phi^\pm} &= - \frac{1}{2} \left( \frac{eA}{\hbar}\right)^2\,t \sum_{\bm{\tau}} (\bm{\tau}.\bm{e})^2 \Phi_{\bm{\tau}}^\pm  
\label{A2}\\
\sum_{\bm{\tau}} (\bm{\tau}.\bm{e})^2  \braket{\bm{\tau}}{\Phi_+}&=\textbf{e}.\bar{\bar{\textbf{q}}}.\textbf{e} =
\frac{3}{8}a_{bn}^2 (e_x-ie_y)^2 |\Phi_{\bm{\tau}}^\pm |\; ,
\end{align}
where $\bar{\bar{\textbf{q}}}_{\Phi^\pm} = \sum_{\bm{\tau}} (\bm{\tau} \otimes \bm{\tau}) \Phi_{\bm{\tau}}^\pm$ is a quadrupolar matrix element.

The last result is remarkable. It shows first that the matrix element is similar to that of the first order term and therefore that the corresponding two-photon process should be strongly visible. Then we see that the favored circular polarization for the two-photon process is the opposite of that of the one-photon process. Finally this direct contribution to first order in $A^2$ is  stronger than the contribution discussed above coming from the second order perturbation theory. The latter one is of order $t^2/\Delta^2$ less than the former one. It is interesting also to analyze what happens in reciprocal space. Then the Peierls substitution amounts to replace $\textbf{k}$ by  
$\textbf{k}-e\textbf{A}/\hbar$ and the non-vanishing $A^2$ contributions appear only if $f(\textbf{k})$ is expanded up to $k^2$ terms. In the context of studies on graphene, this corresponds to including so-called warping effects. In the context of $\textbf{k}.\textbf{p}$ methods such effects would also appear if coupling with other conduction and valence bands are included.
\footnote{M. Glazov, private communication; this is also discussed in Ref.[\onlinecite{Glazov2017}] for the case of TMD.}

\subsubsection{One- and two-photon absorption in bulk \hbn}
\label{stacking}

As mentioned previously and, as far as the band structure is concerned, the extension to three dimensional stackings of  BN layers is fairly simple. 
On the other hand the excitonic formalism should also be extended to the case where there are several atoms per unit cell. In practice, if we continue to define exciton states using the separation between electrons and holes, we must add labels indicating in which type of plane they are.  The general corresponding TB formalism is described elsewhere,\cite{Paleari2018} but if we are principally interested in the ground state excitons, the discussion becomes simpler. Actually the $1s$ exciton of the monolayer discussed above remains confined within a single plane in  bulk \hbn\  if the hole is fixed in this plane.\cite {Arnaud2006}  Along the $0z$ stacking direction we have therefore to treat a problem similar to that of a Frenkel exciton, where the 2D exciton plays the part of an atomic excitation. We can then build two different excitonic Bloch functions along the $0z$ stacking direction which we call $\ket{\Phi_A}$ and $\ket{\Phi_{A'}}$. Introducing interplanar transfer integrals in the TB model couples these two states, and since there is an inversion center between any pair of $A$ and $A'$ planes the final exciton eigenstates should be the usual bonding and antibonding states $\ket{\Phi_A} \pm  \ket{\Phi_{A'}}$. The $1s$ state of the monolayer becomes two split states (Davydov splitting) separated by a small energy (see also Ref.[\onlinecite{Koskelo2017,Paleari2018}]).    \ai\ calculations show that the bonding state is the ground state. Since the threefold symmetry is preserved, the two states are still themselves twofold degenerate.

Let us now look at the one-photon absorption process. We still consider a polarization $\textbf{e}$ parallel to the planes. The total  $\textbf{d}_\Phi $  dipole for the $AA'$ stacking is  therefore equal to $\textbf{d}_{\Phi_A} \pm \textbf{d}_{\Phi_{A'}} $. On the other due to the inversion symmetry $\textbf{d}_{\Phi_{A'}}  = -\textbf{d}_{\Phi_A}$ and finally the dipole vanishes for the bonding state whereas the upper anti-bonding state is bright.
In the case of the two-photon process, we can use a similar argument, but since now $\bar{\bar{\textbf{q}}}_{\Phi_{A'}}   =  \bar{\bar{\textbf{q}}}_{\Phi_{A}} $, the situation is reversed: the bright exciton is the lower bonding one. 


\section{Results and discussion}
\label{sec:res}
\subsection{\ai\ calculations}
\label{sec:resab}

\begin{figure}[t]
\centering
\includegraphics[width=0.48\textwidth]{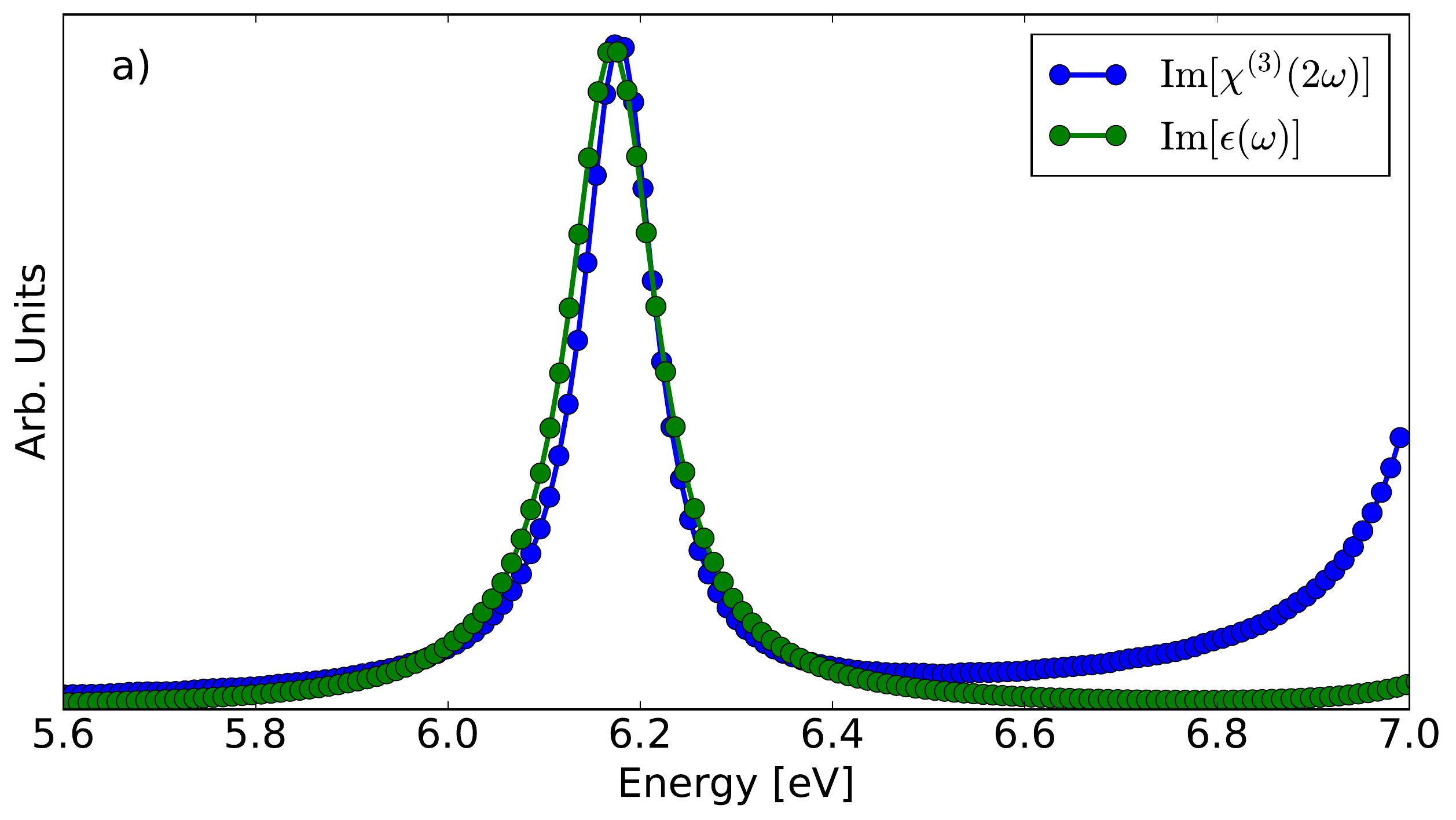}
\includegraphics[width=0.48\textwidth]{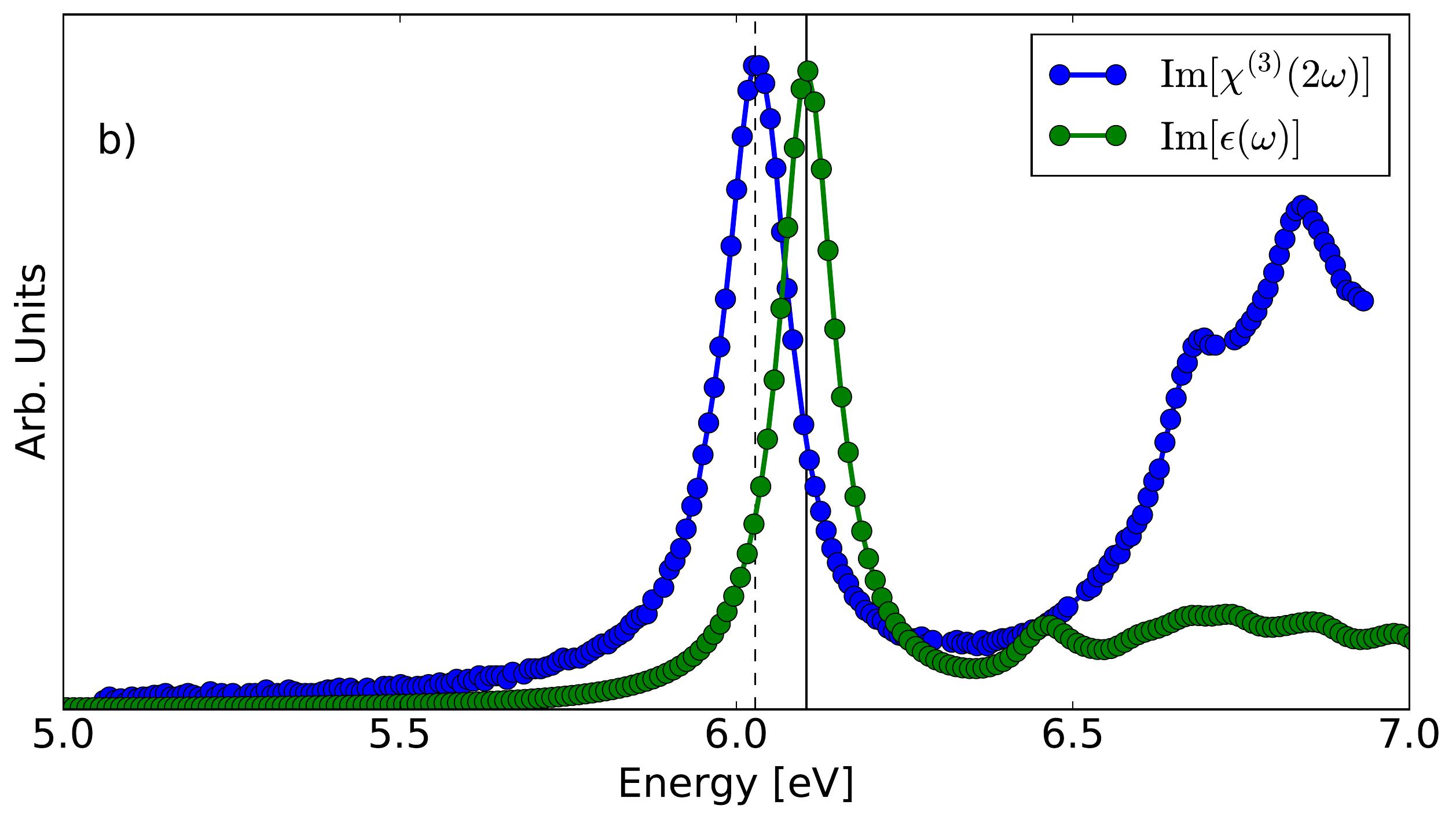}
    \caption{\footnotesize{Two-photon absorption and imaginary part of the dielectric constant in single layer (panel $a$) and bulk \hbn\ (panel $b$). The two curves have been rescaled in such a way to have the same intensity at the maximum position. Vertical lines in panel $(b)$ indicates the position of the maximum.} 
\label{tpa_vs_bulk}}
\end{figure}
All operators in Eq.~\eqref{tdbse_shf} and Eq.~\eqref{eq:hmb} are expanded in the basis set of the \ks\ band states which can be obtained from a standard DFT code. Specifically, we used the \textsc{Quantum ESPRESSO} code\cite{Quantum_espresso2009} where the wavefunctions are expanded in plane waves with a cutoff of $60$~Ry and the effect of core electrons is simulated by norm-conserving pseudopotentials.  A $ 12 \times 12 \times 4$ ($ 12 \times 12 \times 1$ for the single layer) $\textbf{k}$-point shifted grid has been used to converge the electronic density. The band states are obtained from the diagonalization of the \ks\  eigensystem. 
In order to simulate an isolated \hbn\ layer we used a supercell approach with a layer-layer distance of 20 a.u. in the perpendicular direction. 
The scissor operator entering Eq.~\eqref{eq:hmb} is chosen so as to reproduce the position of the first bright excitons in the absorption spectra of bulk and monolayer \hbn\ from Ref.[\onlinecite{Wirtz2006}].
We expanded $| v_{m\textbf{k}} \rangle$ in terms of \ks\ eigenstates and we evolved the coefficients of the bands between 2 and 7 in the monolayer (7 and 12 in the bulk) in Eq.~\eqref{tdbse_shf}. We used a $ 15 \times 15 \times 5$ ($ 12 \times 12 \times 1$ in the single layer) $\textbf{k}$-points $\Gamma$-centered sampling in the real-time simulations which guarantee the convergence of the first peak in the spectra.

Equation~\eqref{tdbse_shf} is solved numerically\footnote{The approach is implemented in the \textsc{Lumen} extension of the \textsc{Yambo} many-body code.~\cite{YAMBO2009} The source code is available from \texttt{http://www.attaccalite.com/lumen/}} for a time interval of $120$~fs using the numerical approach described in Ref.~[\onlinecite{Souza2004}](originally taken from Ref.~[\onlinecite{Koonin1990}]) with a time step of $\Delta t = 0.01$~fs, which guarantees for numerically stable and sufficiently accurate simulations. A dephasing term corresponding to  a finite broadening of about 0.05 eV is introduce in order to simulate the experimental broadening.\cite{Attaccalite2013} 

In Figure~\ref{tpa_vs_bulk}, panel $(a)$, the two-photon resonant third-order susceptibility at $\chi^{(3)}_{yyyy}(-\omega; \omega, \omega, -\omega)$ proportional to the TPA---is compared with the imaginary part of the dielectric constant $\epsilon_2(\omega)$ for the monolayer \hbn. To facilitate the comparison there is a factor $2$ between the energy scale of the spectra. The first peak of the TPA is found exactly at half the photon energy of the first peak of the imaginary part of the dielectric constant. That means that the two- and one-photon absorption are resonant with the same exciton.
In panel $(b)$ the same comparison is shown for bulk \hbn. In this case the first peak of the TPA is found 0.076~eV below half the photon energy of the first peak of $\epsilon_2$. That means that the two-photon absorption is resonant with an exciton at lower energy than the one of one-photon absorption. We also obtained $\epsilon_2(\omega)$ by solving the standard $GW$+Bethe-Salpeter equation~\cite{YAMBO2009} and diagonalized the excitonic two-particle Hamiltonian. We found---in agreement with previous studies~\cite{Wirtz2008} that the lowest exciton in the linear optical response of bulk \hbn\ is indeed dark. The position of this dark exciton is consistent with the splitting deduced from the TPA calculations.
The \ai\ results are then fully consistent with the discussion in Sec.~\ref{stacking}. In the monolayer the ground state exciton is visible in both one-photon and two photon absorption.  Instead, in bulk $AA'$ BN, the lowest exciton (pair) is dark in linear optics but becomes visible in two photon absorption. In fact, as explained in Sec.~\ref{stacking}, the two lowest exciton pairs in the bulk are due to the bonding and anti-bonding combination of excitons in each layer and thus obey  different selection rules. 

Finally, broader features at higher energies are visible in the spectra. Those features are known to originate from both in-plane excitons with different symmetries\cite{Galvani2016} and from interplanar excitations. Notice that those peaks may not be fully converged.~\footnote{Excitons at higher energy are more difficult to characterize because they require a denser $k$-point sampling. As specified above the $k$-point samplings are chosen to guarantee the convergence of the first peak in the spectra that are the focus of this study.}

\subsection{Tight-binding calculations}
\label{sec:restb}

\subsubsection{Monolayer}
\begin{figure}[t]
\centering
\includegraphics[width=0.5\textwidth]{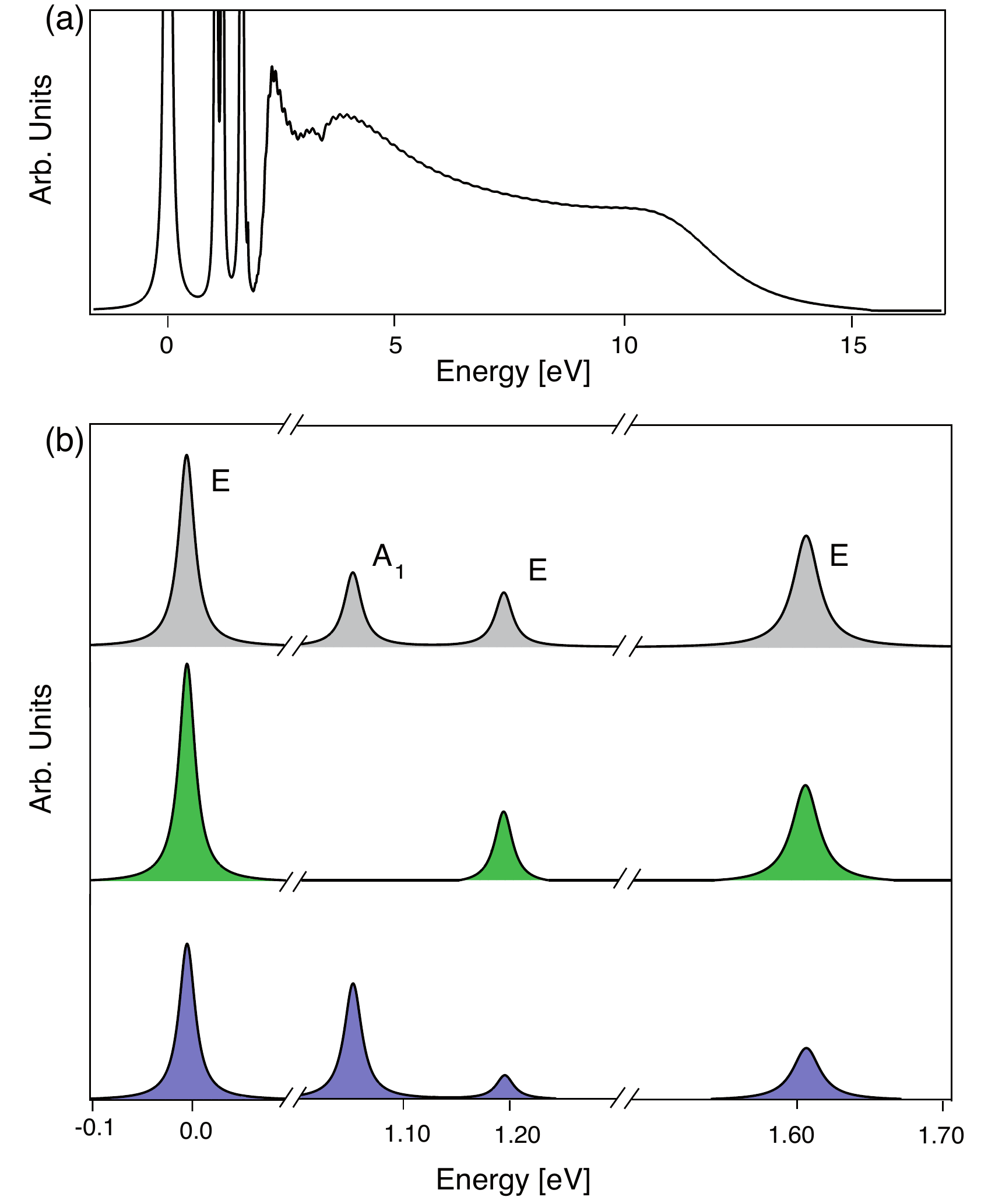}
\caption{\footnotesize{a) One photon spectrum of the 2D BN monolayer with excitonic effects, but without taking into account the exact selection rules (see the main text); b) zoom in the exciton region; top: general spectrum; middle: one-photon optical spectrum; bottom: two-photon optical spectrum. For convenience a broadening of $10^{-2}$ eV  has been applied, via the imaginary part of $z$ in the Green functions. The $A_1$ exciton is only dark in the one-photon spectrum. The origin of energy is taken at the position of the lowest exciton, which can also be labelled as a $1s$ exciton whereas the other ones derive from $2s$ and $2p$ states. Full \textit{ab initio} calculations including exchange effects predict the $A_1$ level to be at a higher energy.\cite{Galvani2016}}}
\label{spectre1ph_2D}
\end{figure}

In the case of the monolayer we have used the TB hamiltonian (\ref{Heh}) with the parameters determined in Ref.[\onlinecite{Galvani2016}] and have calculated the one-photon absorption spectrum from the Green function  $ \bra{I} G(z) \ket{I}, \ket{I} = \sum_{\bm{\tau}}(\bm{\tau}.\textbf{e})\ket{\tau}$, 
which is actually independent of the choice for $\textbf{e}$. This is conveniently done using the recursion method. A cluster of $4.10^4$ atoms is used and  $100$ recursion levels are calculated. The spectrum, proportional to the imaginary part of the optical dielectric constant is also proportional to the imaginary part of the Green function. We have first calculated the spectrum in the absence of excitonic effects ($U=0$ in the hamiltonian). It is actually quite similar to the density of states of the triangular lattice.\cite{Galvani2016} In the presence of excitonic effects the excitons appear as bound states below the continuum. We have checked that they appear at the same positions as those determined from a full diagonalization of the hamiltonian.\cite{Galvani2016} Furthermore by comparing the optical spectrum to that obtained from a Green function matrix element without any particular symmetry (here $\bra{\bm{\tau}} G \ket{\bm{\tau}}$) so that all excitons have a finite weight, we can check the selection rules: the non degenerate exciton of symmetry $A_1$ is actually dark in the optical response (Fig.~ \ref{spectre1ph_2D}).

In the case of two-photon absorption, the calculation of the main contribution due to $H_{I2}$,  the quadratic term in $A$, can be calculated in a similar way; it suffices to modify the matrix element of the Green function, by taking as initial vector $\ket{I'} = \sum_{\bm{\tau}}(\bm{\tau}.\textbf{e})^2\ket{\tau}$. The ground state exciton is then found to be bright  as expected.

\subsubsection{Bulk AA'}

\begin{figure}[t]
\centering
\includegraphics[width=0.45\textwidth]{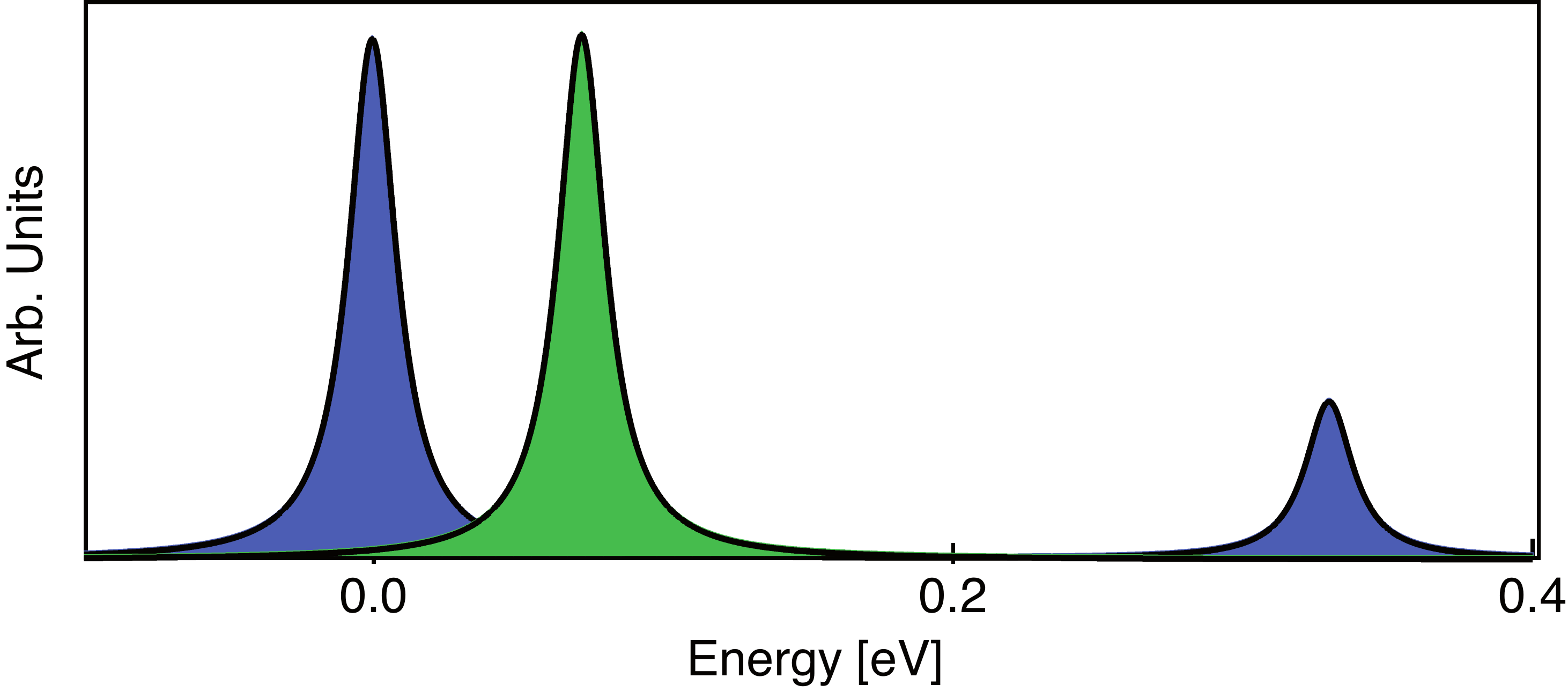}
\caption{\footnotesize{One- (green) and two-photon (blue) absorption spectrum of AA'  \hbn. Compare with Fig.\ \ref{tpa_vs_bulk}b}.}
\label{spectreAA'}
\end{figure}

We have generalized the TB formalism to the 3D $AA'$ stacking. Once the appropriate hamiltonian is defined the recursion method can again be used. Here, the cluster considered contains $8.10^4$ atoms and $100$ recursion levels are calculated. According to our previous discussion, to study the two split ground state excitons we have now to use  starting states $\ket{I}$ which are bonding and antibonding combinations of excitons in $A$ and $A'$ planes . The results are shown in Figs.~\ref{spectreAA'}. As predicted in the previous section we do find that the lower bonding state is dark for one-photon absorption and bright for two-photon absorption, and conversely for the upper antibonding state.

\subsection{Selection rules}
\label{sec:rules}

Let us summarize the selection rules which apply to \hbn\  as well as to TMD.

\subsubsection{One-photon selection rules}

We consider first the case of the monolayer, and recall  the usual rules for Wannier-Mott excitons within the continuous hydrogenic model. The exciton wave function is written in the form $\Phi(\textbf{r}_h,\textbf{r}_e)=\phi_{\textbf{k}_0c}(\textbf{r}_e)\phi_{\textbf{k}_0v}(\textbf{r}_h)g(\textbf{r}_e-\textbf{r}_h)$, where the $\phi_{\textbf{k}_0}$ are the single particle Bloch functions at point $\textbf{k}_0$ corresponding to the considered direct gap, and $g(\textbf{r})$, the envelope function,  is the solution of the hydrogenic-like excitonic equation for the relative coordinate $\textbf{r}=\textbf{r}_e-\textbf{r}_h$. In our case $\textbf{k}_0$ is a $K$ or $K'$ point. Then the optical matrix element is the standard one-particle matrix element at this point weighted by $g(\textbf{r}=0)$. The full excitonic symmetry is the symmetry of the above product, but notations generally use the symmetry of $g(\textbf{r})$. Optical one-photon transitions are then only allowed for $s$ states. On the other hand the direct optical transition is allowed since the standard matrix element varies as $v_F (e_x\pm ie_y)$ (see Eq.(\ref{matrix_element})), corresponding to polarisations $\sigma_\pm$, depending on the valley $K$ or $K'$. Hence the rule for the exciton one-photon transitions: only $s$ states are allowed and the optical transitions involve $\sigma_\pm$ polarisations. These rules are weakened if the dependence on $\textbf{k}$ of the matrix elements is taken into account.\cite{Gong2017}  At higher values of $\textbf{k}$, deviations from an isotropic hamiltonian appear, with so-called warping effects, and the usual orbital selection rules are modified.\cite{saynatjoki2017ultra} Furthermore, when the exciton size decreases, its size in reciprocal space increases and intervalley coupling is no longer negligible. Although $\textbf{k}.\textbf{p}$ method can be used,\cite{Glazov2017} it is then much more convenient to work in real space so as to take the triangular symmetry fully in account.\cite{Galvani2016} 

The excitonic states characterized by the wave function $\Phi_{\textbf{R}} = \braket{\textbf{R}}{\Phi^\pm}$ for the monolayer can then be classified according to the representations of the $C_{3v}$ point group. Among the three representations $A_1, A_2$ and $E$, only the two-dimensional  representation $E$ is optically active, as discussed in Section \ref{opt}. Let us precise the correspondence between the continuous description in  terms of $s,p, \dots$ states characterizing the symmetry of the envelope function and the present description using the discrete crystal symmetry. For that,  we have to include the symmetry of the product of $\phi_{\textbf{k}_0c}(\textbf{r}_e)\phi_{\textbf{k}_0v}(\textbf{r}_h)$ at $K$ point. The relevant  group of vector $\textbf{K}$ is the $C_3$ point group so that in our case this product is multiplied by $e^{\pm 2i\pi/3}$ under a rotation of $\pm 2\pi/3$. In other words it transforms as $e^{i\varphi}$ if $\varphi$ is the azimuthal angle. For a level of symmetry characterized by an angular momentum $m$, the envelope function of the corresponding exciton states varies as $e^{\pm im\varphi}$.

When $m\neq 0$ the level is twofold degenerate and the full wave function varies as $e^{i(1 \pm m)\varphi}$. The same is true at $K'$ point provided $\varphi$ is changed into $-\varphi$. So finally the level shows a fourfold degeneracy. This is not consistent with representation group theory which says that the degeneracy is at most equal to two in general. The crystal symmetry is lower than the continuous symmetry and the degeneracy is lifted. Consider the $p$ states for example ($m=\pm 1$). It has been shown in Ref.[\onlinecite{Galvani2016}] that the four $p$ levels decompose according to the $E+A_1+A_2$ representations. This can be analysed in a first step as an intravalley splitting into distinct $1\pm 1$ states. Within the continuous model this can be attributed to Berry phase effects.\cite{Zhou2015,Srivastava2015} Another effect of the lower crystalline symmetry is that $m_{\text{tot}} = 1 \pm m$ should be counted modulo three, since only rotations of order three are involved. Thus only states $m_{\text{tot}} = 0,\pm 1$ are relevant in each valley. Here the allowed values in valley $K$ are 0 and 2, \textit{i.e.} 0 and -1, modulo 3. In valley $K'$ the allowed values are 0 and +1. These rules are  fairly well-known\cite{Wu2015,Xiao2015,Galvani2016,Glazov2017,Gong2017} but have been recently re-discovered and discussed in detail.\cite{Cao2018,Zhang2018}

Then, if intervalley coupling is accounted for, the $\pm 1$ states combine to produce a global $E$ symmetry, whereas the $m_{\text{tot}} = 0$ states combine to form $A_1$ and $A_2$ states. Now, only $E$ states are one-photon bright, since the velocity also transforms as $E$. More precisely, its components $\textbf{v}.(\hat{\textbf{x}}\pm i \hat{\textbf{y}})$ transform as $e^{\pm i\varphi}$. The ground state $1s$ exciton is obviously bright and form an $E$ state with a very strong oscillator strength, but we see that the $2p$ states give rise also to a bright exciton. This is really a ``modulo 3 '' effect which transforms a forbidden $m_{\text{tot}}=2$ state into an allowed $m_{\text{tot}}=-1$ state. Notice also that for a given valley the circular polarizations for the  $s$ and  $p$ states are opposite. 

Consider now the bulk $AA'$ stacking. We still assume the light polarization to be within the planes. Then according to the discussion given in Section \ref{stacking}, we  have just to consider a linear superposition of monolayer $E$ states. For this stacking this gives rise to Davydov pairs of bonding and antibonding states, and only antibonding states are bright. Interlayer excitons where the hole and the electron are in different planes can be discussed as well and are relevant if the light polarization is along the stacking axis. 

There are therefore two types of selection rules for such lamellar structures for a polarization in the planes. The first one ensures the existence of a dipole within the planes (exciton of $E$ symmetry). The second depends on the constructive or destructive arrangement of the dipoles in the stacking.

\subsubsection{Two photon selection rules}

\begin{figure}[t]
\centering
\includegraphics[width=0.50\textwidth]{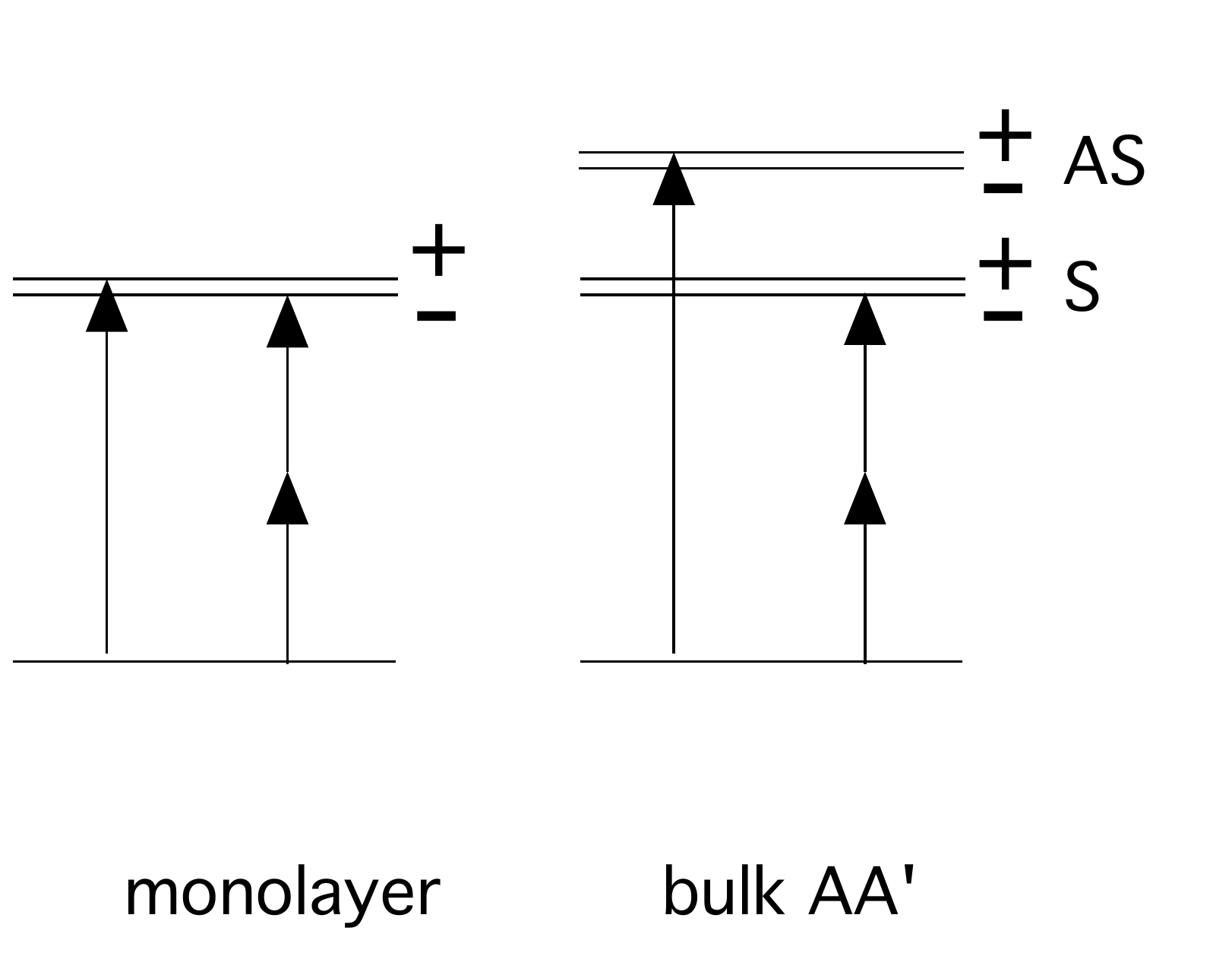}
\caption{\footnotesize{Allowed transitions towards the lowest excitonic states with circularly polarized light. Left: Monolayer; the $\Phi^+ (\Phi^-)$ is excited with a one-photon (two-photon) process, where $ \Phi^\pm$ denote the two circularly components of the degenerate $1s$ exciton. Right: Bulk $AA'$ stacking; there are now two  antisymmetric (AS, odd, one-photon allowed) and symmetric (S, even, two-photon allowed) degenerate states, separated by a Davydov splitting.}}
\label{transitions}
\end{figure}

Within the continuous model, the general statement concerning selection rules is that only $p$ states are visible, corresponding to $\nabla g(\textbf{r})  \neq 0$.\cite{Berkelbach2015}  What is changing here is the symmetry of the coupling with light which has now a tensorial character and varies therefore as $e^{\pm i\varphi}e^{\pm i\varphi}$ for the monolayer. If the rule of $m$ modulo 3   is added, this means that  all $m_{\text{tot}}$ values are allowed. In the discrete case, it varies as $E \otimes E = E+A_1+A_2$ which indicates also that all excitons are in principle bright. We have seen in particular that the oscillator strength for the ground state $1s$ exciton is found actually to be very strong. In the case of the bulk $AA'$ stacking, the bright excitons in the Davydov pairs are the bonding states. This is a familiar selection rule: In the presence of a symmetry centre odd (even) states are one(two)-photon allowed. Then the difference with the one-photon selection rules is less important than in the usual continuous model, but at least in the case of the $AA'$ stacking combining both processes can be used to discriminate between the components of the Davydov doublets (Fig.\ref{transitions}). The two-photon selection rules for the monolayer agree with those derived by Xiao et al.\cite{Xiao2015} who, however, do not discuss the possibility of the splitting of $2p$ states. Since they consider only circular polarizations, they do not discuss either the possibility of exciting $m=0$ (or $A_1, A_2$) states, but the corresponding oscillator strengths should be weak since they imply intervalley interactions.

\subsubsection{Experimental results}

One of our main result is the prediction that the ground state $1s$ exciton which is bright in one-photon absorption should be bright also in two-photon processes, with opposite circular polarizations. The best  experimental evidence is certainly the observation in TMD of resonant second harmonic generation (SHG).
\footnote{Non resonant SHG experiments in \hbn\  are described in [\onlinecite{Kim2013,Li2013}] whereas \ai\ calculations are presented in [\onlinecite{Gruning2014}].}
In the absence of symmetry center, SHG is allowed for a dichalcogenide layer and it is actually found to be strongly resonant when the $1s$ level is excited in a two-photon process. Furthermore the circular polarization of the  $2\omega$ emission is actually opposite to that of the excitation.\cite{Wang2015,Xiao2015} In the case of \hbn\  only two-photon photoluminescence excitation (PLE) spectra are available for the bulk phase.\cite{Cassabois2016} They do show a peak slightly above the one-photon main peak whereas we predict a peak below it. The situation is  complicated by the fact that the gap is indirect. This is important for the interpretation of luminescence spectra but it is suspected that absorption spectra are governed by direct transitions.\cite{Schue2018} The difference between one-photon and two-photon spectra has been interpreted as the signature of $s$ and $p$ states respectively. The present analysis show that this is probably not true because of the non conventional selection rules. Another argument is that the $2p$ and $2s$ are predicted to be well above the $1s$ states (about 1 eV for the monolayer, 0.5 eV for the bulk) and cannot therefore be involved in the observed splitting. But the precise interpretation of the observed splitting remains to be correlated with the calculated Davydov splitting.

To summarize we have performed  tight-binding and \ai\ calculations of two-photon absorption in monolayer and bulk boron nitride and found that at low energy it is driven by excitonic effects. The ground state $1s$ exciton is predominant as for one-photon absorption, indicating strong deviations from the usual selection rules which are explained within a simple TB model taking into account the crystalline symmetry.

\section*{acknowledgments}
The French National Agency for Research (ANR) is acknowledged for funding this work under the project GoBN (Graphene on Boron Nitride Technology), Grant No. ANR-14-CE08-0018. The research leading to these results has  received funding from the European Union Seventh Framework Program under grant agreement no. 696656 GrapheneCore1. L. Sponza and J. Barjon are gratefully acknowledged for fruitful discussions.

\appendix*
\section{On the equivalence of length and velocity gauge in tight-binding}
\label{gauge_equivalence}
Optical properties are usually treated using the so-called velocity gauge: $\textbf{p}$ is replaced by  $\textbf{p}-e\textbf{A}$. As discussed in the main text this implies that to calculate the response to second order in $\textbf{A}$ we have to calculate a first order perturbation term in $A^2$ and a second order perturbation term with respect to  $\textbf{A}$. In the length gauge where the interaction term in the hamiltonian is equal to $e\,\textbf{r}.\bm{\mathcal{E}}$, there is no quadratic term. Since approximations are made, it is useful to check gauge invariance. If we refer to the discussion made in the velocity gauge in Section \ref{2nd_order} we have therefore just to calculate the second order perturbation term proportional to $\Delta\bra{\Phi}(\textbf{r}.\textbf{e})(\textbf{r}.\textbf{e})\ket{\emptyset}$. We have used the fact that $\mathcal{\bm{E}}= i\omega\textbf{A}$ and that $\omega \simeq \Delta$. The first $\textbf{r}$ generates electron-hole pairs $ \textbf{r} _{\textbf{n}\textbf{m}}  a_{\textbf{n}_c}^+ a_{\textbf{m}_v}^{} \ket{\emptyset}$. To lowest order in $t/\Delta, \textbf{r} _{\textbf{n}\textbf{m}} $ vanishes, and we must use the improved Wannier basis defined in Eqs (\ref{Wannier1}-\ref{Wannier2}), and then $\textbf{r} _{\textbf{n}\textbf{m}}  \simeq  -{(t/2\Delta)}(\textbf{n} -\textbf{m})  $, where $(\textbf{n} -\textbf{m})$ is a first neighbour vector $\bm{\tau}$, so that:
\begin{equation*}
\textbf{r} \ket{\emptyset} \simeq  - \frac{t}{2\Delta} \sum_{\bm{\tau}} \bm{\tau}  \ket{\bm{\tau}}
\end{equation*}
 The second $\textbf{r}$ operator  connects intraband conduction and valence states, and to lowest order in $t/\Delta$ it can be checked that 
$\textbf{r} \ket{\bm{\tau}} \simeq \bm{\tau} \ket{\bm{\tau}}$, so that, finally:
\begin{equation*}
\Delta\bra{\Phi} (\textbf{r}.\textbf{e})(\textbf{r}.\textbf{e})\ket{\emptyset} \simeq - \frac{t}{2} \sum_{\bm{\tau}} (\bm{\tau}.\bm{e})^2 \Phi_{\bm{\tau}}^\pm  \; ,
\end{equation*}
which is exactly the result derived in Eq.(\ref{A2}). Thus to lowest order, the second order perturbation theory in the length gauge reproduces the first order term as a function of $(\textbf{A}.\textbf{e})^2$ in the velocity gauge. The general equivalence of both gauges for non-linear responses of higher order has also been discussed recently in Ref. [\onlinecite{Ventura2017}].

\bibliography{tpa}

\begin{thebibliography}{72}%
\makeatletter
\providecommand \@ifxundefined [1]{%
 \@ifx{#1\undefined}
}%
\providecommand \@ifnum [1]{%
 \ifnum #1\expandafter \@firstoftwo
 \else \expandafter \@secondoftwo
 \fi
}%
\providecommand \@ifx [1]{%
 \ifx #1\expandafter \@firstoftwo
 \else \expandafter \@secondoftwo
 \fi
}%
\providecommand \natexlab [1]{#1}%
\providecommand \enquote  [1]{``#1''}%
\providecommand \bibnamefont  [1]{#1}%
\providecommand \bibfnamefont [1]{#1}%
\providecommand \citenamefont [1]{#1}%
\providecommand \href@noop [0]{\@secondoftwo}%
\providecommand \href [0]{\begingroup \@sanitize@url \@href}%
\providecommand \@href[1]{\@@startlink{#1}\@@href}%
\providecommand \@@href[1]{\endgroup#1\@@endlink}%
\providecommand \@sanitize@url [0]{\catcode `\\12\catcode `\$12\catcode
  `\&12\catcode `\#12\catcode `\^12\catcode `\_12\catcode `\%12\relax}%
\providecommand \@@startlink[1]{}%
\providecommand \@@endlink[0]{}%
\providecommand \url  [0]{\begingroup\@sanitize@url \@url }%
\providecommand \@url [1]{\endgroup\@href {#1}{\urlprefix }}%
\providecommand \urlprefix  [0]{URL }%
\providecommand \Eprint [0]{\href }%
\providecommand \doibase [0]{http://dx.doi.org/}%
\providecommand \selectlanguage [0]{\@gobble}%
\providecommand \bibinfo  [0]{\@secondoftwo}%
\providecommand \bibfield  [0]{\@secondoftwo}%
\providecommand \translation [1]{[#1]}%
\providecommand \BibitemOpen [0]{}%
\providecommand \bibitemStop [0]{}%
\providecommand \bibitemNoStop [0]{.\EOS\space}%
\providecommand \EOS [0]{\spacefactor3000\relax}%
\providecommand \BibitemShut  [1]{\csname bibitem#1\endcsname}%
\let\auto@bib@innerbib\@empty
\bibitem [{\citenamefont {Callis}(1997)}]{Callis1997}%
  \BibitemOpen
  \bibfield  {author} {\bibinfo {author} {\bibfnamefont {P.~R.}\ \bibnamefont
  {Callis}},\ }\bibfield  {title} {\enquote {\bibinfo {title}
  {Two-photon--induced fluorescence},}\ }\href@noop {} {\bibfield  {journal}
  {\bibinfo  {journal} {Annual Review of Physical Chemistry}\ }\textbf
  {\bibinfo {volume} {48}},\ \bibinfo {pages} {271--297} (\bibinfo {year}
  {1997})}\BibitemShut {NoStop}%
\bibitem [{\citenamefont {Shen}(1984)}]{Shen1984}%
  \BibitemOpen
  \bibfield  {author} {\bibinfo {author} {\bibfnamefont {Yuen-Ron}\
  \bibnamefont {Shen}},\ }\bibfield  {title} {\enquote {\bibinfo {title} {The
  principles of nonlinear optics},}\ }\href@noop {} {\bibfield  {journal}
  {\bibinfo  {journal} {New York, Wiley-Interscience, 1984, 575 p.}\ }
  (\bibinfo {year} {1984})}\BibitemShut {NoStop}%
\bibitem [{\citenamefont {Wang}\ \emph {et~al.}(2005)\citenamefont {Wang},
  \citenamefont {Dukovic}, \citenamefont {Brus},\ and\ \citenamefont
  {Heinz}}]{Wang2005}%
  \BibitemOpen
  \bibfield  {author} {\bibinfo {author} {\bibfnamefont {F.}~\bibnamefont
  {Wang}}, \bibinfo {author} {\bibfnamefont {G.}~\bibnamefont {Dukovic}},
  \bibinfo {author} {\bibfnamefont {L.~E.}\ \bibnamefont {Brus}}, \ and\
  \bibinfo {author} {\bibfnamefont {T.~F.}\ \bibnamefont {Heinz}},\ }\bibfield
  {title} {\enquote {\bibinfo {title} {The optical resonances in carbon
  nanotubes arise from excitons},}\ }\href@noop {} {\bibfield  {journal}
  {\bibinfo  {journal} {Science}\ }\textbf {\bibinfo {volume} {308}},\ \bibinfo
  {pages} {838--841} (\bibinfo {year} {2005})}\BibitemShut {NoStop}%
\bibitem [{\citenamefont {Cassabois}\ \emph {et~al.}(2016)\citenamefont
  {Cassabois}, \citenamefont {Valvin},\ and\ \citenamefont
  {Gil}}]{Cassabois2016}%
  \BibitemOpen
  \bibfield  {author} {\bibinfo {author} {\bibfnamefont {Guillaume}\
  \bibnamefont {Cassabois}}, \bibinfo {author} {\bibfnamefont {Pierre}\
  \bibnamefont {Valvin}}, \ and\ \bibinfo {author} {\bibfnamefont {Bernard}\
  \bibnamefont {Gil}},\ }\bibfield  {title} {\enquote {\bibinfo {title}
  {Hexagonal boron nitride is an indirect bandgap semiconductor},}\ }\href@noop
  {} {\bibfield  {journal} {\bibinfo  {journal} {Nature Photonics}\ }\textbf
  {\bibinfo {volume} {10}},\ \bibinfo {pages} {nphoton--2015} (\bibinfo {year}
  {2016})}\BibitemShut {NoStop}%
\bibitem [{\citenamefont {Barros}\ \emph {et~al.}(2006)\citenamefont {Barros},
  \citenamefont {Capaz}, \citenamefont {Jorio}, \citenamefont {Samsonidze},
  \citenamefont {Souza~Filho}, \citenamefont {Ismail-Beigi}, \citenamefont
  {Spataru}, \citenamefont {Louie}, \citenamefont {Dresselhaus},\ and\
  \citenamefont {Dresselhaus}}]{Barros2006}%
  \BibitemOpen
  \bibfield  {author} {\bibinfo {author} {\bibfnamefont {E.~B.}\ \bibnamefont
  {Barros}}, \bibinfo {author} {\bibfnamefont {R.~B.}\ \bibnamefont {Capaz}},
  \bibinfo {author} {\bibfnamefont {A.}~\bibnamefont {Jorio}}, \bibinfo
  {author} {\bibfnamefont {G.~G.}\ \bibnamefont {Samsonidze}}, \bibinfo
  {author} {\bibfnamefont {A.~G.}\ \bibnamefont {Souza~Filho}}, \bibinfo
  {author} {\bibfnamefont {S.}~\bibnamefont {Ismail-Beigi}}, \bibinfo {author}
  {\bibfnamefont {C.~D.}\ \bibnamefont {Spataru}}, \bibinfo {author}
  {\bibfnamefont {S.~G.}\ \bibnamefont {Louie}}, \bibinfo {author}
  {\bibfnamefont {G.}~\bibnamefont {Dresselhaus}}, \ and\ \bibinfo {author}
  {\bibfnamefont {M.~S.}\ \bibnamefont {Dresselhaus}},\ }\bibfield  {title}
  {\enquote {\bibinfo {title} {Selection rules for one-and two-photon
  absorption by excitons in carbon nanotubes},}\ }\href@noop {} {\bibfield
  {journal} {\bibinfo  {journal} {Physical Review B}\ }\textbf {\bibinfo
  {volume} {73}},\ \bibinfo {pages} {241406} (\bibinfo {year}
  {2006})}\BibitemShut {NoStop}%
\bibitem [{\citenamefont {Li}\ \emph {et~al.}(2015)\citenamefont {Li},
  \citenamefont {Dong}, \citenamefont {Zhang}, \citenamefont {Zhang},
  \citenamefont {Feng}, \citenamefont {Wang}, \citenamefont {Zhang},\ and\
  \citenamefont {Wang}}]{li2015giant}%
  \BibitemOpen
  \bibfield  {author} {\bibinfo {author} {\bibfnamefont {Yuanxin}\ \bibnamefont
  {Li}}, \bibinfo {author} {\bibfnamefont {Ningning}\ \bibnamefont {Dong}},
  \bibinfo {author} {\bibfnamefont {Saifeng}\ \bibnamefont {Zhang}}, \bibinfo
  {author} {\bibfnamefont {Xiaoyan}\ \bibnamefont {Zhang}}, \bibinfo {author}
  {\bibfnamefont {Yanyan}\ \bibnamefont {Feng}}, \bibinfo {author}
  {\bibfnamefont {Kangpeng}\ \bibnamefont {Wang}}, \bibinfo {author}
  {\bibfnamefont {Long}\ \bibnamefont {Zhang}}, \ and\ \bibinfo {author}
  {\bibfnamefont {Jun}\ \bibnamefont {Wang}},\ }\bibfield  {title} {\enquote
  {\bibinfo {title} {Giant two-photon absorption in monolayer mos2},}\
  }\href@noop {} {\bibfield  {journal} {\bibinfo  {journal} {Laser \& Photonics
  Reviews}\ }\textbf {\bibinfo {volume} {9}},\ \bibinfo {pages} {427--434}
  (\bibinfo {year} {2015})}\BibitemShut {NoStop}%
\bibitem [{\citenamefont {Zhang}\ \emph {et~al.}(2015)\citenamefont {Zhang},
  \citenamefont {Dong}, \citenamefont {McEvoy}, \citenamefont {O'Brien},
  \citenamefont {Winters}, \citenamefont {Berner}, \citenamefont {Yim},
  \citenamefont {Li}, \citenamefont {Zhang}, \citenamefont {Chen} \emph
  {et~al.}}]{zhang2015direct}%
  \BibitemOpen
  \bibfield  {author} {\bibinfo {author} {\bibfnamefont {Saifeng}\ \bibnamefont
  {Zhang}}, \bibinfo {author} {\bibfnamefont {Ningning}\ \bibnamefont {Dong}},
  \bibinfo {author} {\bibfnamefont {Niall}\ \bibnamefont {McEvoy}}, \bibinfo
  {author} {\bibfnamefont {Maria}\ \bibnamefont {O'Brien}}, \bibinfo {author}
  {\bibfnamefont {Sin{\'e}ad}\ \bibnamefont {Winters}}, \bibinfo {author}
  {\bibfnamefont {Nina~C}\ \bibnamefont {Berner}}, \bibinfo {author}
  {\bibfnamefont {Chanyoung}\ \bibnamefont {Yim}}, \bibinfo {author}
  {\bibfnamefont {Yuanxin}\ \bibnamefont {Li}}, \bibinfo {author}
  {\bibfnamefont {Xiaoyan}\ \bibnamefont {Zhang}}, \bibinfo {author}
  {\bibfnamefont {Zhanghai}\ \bibnamefont {Chen}},  \emph {et~al.},\ }\bibfield
   {title} {\enquote {\bibinfo {title} {Direct observation of degenerate
  two-photon absorption and its saturation in ws2 and mos2 monolayer and
  few-layer films},}\ }\href@noop {} {\bibfield  {journal} {\bibinfo  {journal}
  {ACS nano}\ }\textbf {\bibinfo {volume} {9}},\ \bibinfo {pages} {7142--7150}
  (\bibinfo {year} {2015})}\BibitemShut {NoStop}%
\bibitem [{\citenamefont {He}\ \emph {et~al.}(2014)\citenamefont {He},
  \citenamefont {Kumar}, \citenamefont {Zhao}, \citenamefont {Wang},
  \citenamefont {Mak}, \citenamefont {Zhao},\ and\ \citenamefont
  {Shan}}]{he2014tightly}%
  \BibitemOpen
  \bibfield  {author} {\bibinfo {author} {\bibfnamefont {Keliang}\ \bibnamefont
  {He}}, \bibinfo {author} {\bibfnamefont {Nardeep}\ \bibnamefont {Kumar}},
  \bibinfo {author} {\bibfnamefont {Liang}\ \bibnamefont {Zhao}}, \bibinfo
  {author} {\bibfnamefont {Zefang}\ \bibnamefont {Wang}}, \bibinfo {author}
  {\bibfnamefont {Kin~Fai}\ \bibnamefont {Mak}}, \bibinfo {author}
  {\bibfnamefont {Hui}\ \bibnamefont {Zhao}}, \ and\ \bibinfo {author}
  {\bibfnamefont {Jie}\ \bibnamefont {Shan}},\ }\bibfield  {title} {\enquote
  {\bibinfo {title} {Tightly bound excitons in monolayer wse 2},}\ }\href@noop
  {} {\bibfield  {journal} {\bibinfo  {journal} {Physical review letters}\
  }\textbf {\bibinfo {volume} {113}},\ \bibinfo {pages} {026803} (\bibinfo
  {year} {2014})}\BibitemShut {NoStop}%
\bibitem [{\citenamefont {Wang}\ \emph {et~al.}(2017)\citenamefont {Wang},
  \citenamefont {Chernikov}, \citenamefont {Glazov}, \citenamefont {Heinz},
  \citenamefont {Marie}, \citenamefont {Amand},\ and\ \citenamefont
  {Urbaszek}}]{Gang2017}%
  \BibitemOpen
  \bibfield  {author} {\bibinfo {author} {\bibfnamefont {Gang}\ \bibnamefont
  {Wang}}, \bibinfo {author} {\bibfnamefont {A.}~\bibnamefont {Chernikov}},
  \bibinfo {author} {\bibfnamefont {M.~M.}\ \bibnamefont {Glazov}}, \bibinfo
  {author} {\bibfnamefont {T.~F.}\ \bibnamefont {Heinz}}, \bibinfo {author}
  {\bibfnamefont {X.}~\bibnamefont {Marie}}, \bibinfo {author} {\bibfnamefont
  {T.}~\bibnamefont {Amand}}, \ and\ \bibinfo {author} {\bibfnamefont
  {B.}~\bibnamefont {Urbaszek}},\ }\bibfield  {title} {\enquote {\bibinfo
  {title} {Excitons in atomically thin transition metal dichalcogenides},}\
  }\href@noop {} {\bibfield  {journal} {\bibinfo  {journal} {arXiv:1707.05863
  [cond-mat.mtrl-sci]}\ } (\bibinfo {year} {2017})}\BibitemShut {NoStop}%
\bibitem [{\citenamefont {Kang}\ \emph {et~al.}(2016)\citenamefont {Kang},
  \citenamefont {Zhang},\ and\ \citenamefont {Wei}}]{Kang2016}%
  \BibitemOpen
  \bibfield  {author} {\bibinfo {author} {\bibfnamefont {Joongoo}\ \bibnamefont
  {Kang}}, \bibinfo {author} {\bibfnamefont {Lijun}\ \bibnamefont {Zhang}}, \
  and\ \bibinfo {author} {\bibfnamefont {Su-Huai}\ \bibnamefont {Wei}},\
  }\bibfield  {title} {\enquote {\bibinfo {title} {A unified understanding of
  the thickness-dependent bandgap transition in hexagonal two-dimensional
  semiconductors},}\ }\href {\doibase 10.1021/acs.jpclett.5b02687} {\bibfield
  {journal} {\bibinfo  {journal} {The Journal of Physical Chemistry Letters}\
  }\textbf {\bibinfo {volume} {7}},\ \bibinfo {pages} {597--602} (\bibinfo
  {year} {2016})}\BibitemShut {NoStop}%
\bibitem [{\citenamefont {Wu}\ \emph {et~al.}(2015)\citenamefont {Wu},
  \citenamefont {Qu},\ and\ \citenamefont {MacDonald}}]{Wu2015}%
  \BibitemOpen
  \bibfield  {author} {\bibinfo {author} {\bibfnamefont {F.}~\bibnamefont
  {Wu}}, \bibinfo {author} {\bibfnamefont {F.}~\bibnamefont {Qu}}, \ and\
  \bibinfo {author} {\bibfnamefont {A.~H.}\ \bibnamefont {MacDonald}},\
  }\bibfield  {title} {\enquote {\bibinfo {title} {Exciton band structure of
  monolayer ${\mathrm{mos}}_{2}$},}\ }\href {\doibase
  10.1103/PhysRevB.91.075310} {\bibfield  {journal} {\bibinfo  {journal} {Phys.
  Rev. B}\ }\textbf {\bibinfo {volume} {91}},\ \bibinfo {pages} {075310}
  (\bibinfo {year} {2015})}\BibitemShut {NoStop}%
\bibitem [{\citenamefont {Berkelbach}\ \emph {et~al.}(2015)\citenamefont
  {Berkelbach}, \citenamefont {Hybertsen},\ and\ \citenamefont
  {Reichman}}]{Berkelbach2015}%
  \BibitemOpen
  \bibfield  {author} {\bibinfo {author} {\bibfnamefont {T.~C.}\ \bibnamefont
  {Berkelbach}}, \bibinfo {author} {\bibfnamefont {M.~S.}\ \bibnamefont
  {Hybertsen}}, \ and\ \bibinfo {author} {\bibfnamefont {D.~R.}\ \bibnamefont
  {Reichman}},\ }\bibfield  {title} {\enquote {\bibinfo {title} {Bright and
  dark singlet excitons via linear and two-photon spectroscopy in monolayer
  transition-metal dichalcogenides},}\ }\href {\doibase
  10.1103/PhysRevB.92.085413} {\bibfield  {journal} {\bibinfo  {journal} {Phys.
  Rev. B}\ }\textbf {\bibinfo {volume} {92}},\ \bibinfo {pages} {085413}
  (\bibinfo {year} {2015})}\BibitemShut {NoStop}%
\bibitem [{\citenamefont {Xiao}\ \emph {et~al.}(2015)\citenamefont {Xiao},
  \citenamefont {Ye}, \citenamefont {Wang}, \citenamefont {Zhu}, \citenamefont
  {Wang},\ and\ \citenamefont {Zhang}}]{Xiao2015}%
  \BibitemOpen
  \bibfield  {author} {\bibinfo {author} {\bibfnamefont {J.}~\bibnamefont
  {Xiao}}, \bibinfo {author} {\bibfnamefont {Z.}~\bibnamefont {Ye}}, \bibinfo
  {author} {\bibfnamefont {Y.}~\bibnamefont {Wang}}, \bibinfo {author}
  {\bibfnamefont {H.}~\bibnamefont {Zhu}}, \bibinfo {author} {\bibfnamefont
  {Y.}~\bibnamefont {Wang}}, \ and\ \bibinfo {author} {\bibfnamefont
  {X.}~\bibnamefont {Zhang}},\ }\bibfield  {title} {\enquote {\bibinfo {title}
  {{Nonlinear optical selection rule based on valley-exciton locking in
  monolayer ws2}},}\ }\href@noop {} {\bibfield  {journal} {\bibinfo  {journal}
  {Light: Science {\&}Amp; Applications}\ }\textbf {\bibinfo {volume} {4}},\
  \bibinfo {pages} {e366} (\bibinfo {year} {2015})}\BibitemShut {NoStop}%
\bibitem [{\citenamefont {Galvani}\ \emph {et~al.}(2016)\citenamefont
  {Galvani}, \citenamefont {Paleari}, \citenamefont {Miranda}, \citenamefont
  {Molina-S{\'a}nchez}, \citenamefont {Wirtz}, \citenamefont {Latil},
  \citenamefont {Amara},\ and\ \citenamefont {Ducastelle}}]{Galvani2016}%
  \BibitemOpen
  \bibfield  {author} {\bibinfo {author} {\bibfnamefont {T.}~\bibnamefont
  {Galvani}}, \bibinfo {author} {\bibfnamefont {F.}~\bibnamefont {Paleari}},
  \bibinfo {author} {\bibfnamefont {H.~P.~C.}\ \bibnamefont {Miranda}},
  \bibinfo {author} {\bibfnamefont {A.}~\bibnamefont {Molina-S{\'a}nchez}},
  \bibinfo {author} {\bibfnamefont {L.}~\bibnamefont {Wirtz}}, \bibinfo
  {author} {\bibfnamefont {S.}~\bibnamefont {Latil}}, \bibinfo {author}
  {\bibfnamefont {H.}~\bibnamefont {Amara}}, \ and\ \bibinfo {author}
  {\bibfnamefont {F.}~\bibnamefont {Ducastelle}},\ }\bibfield  {title}
  {\enquote {\bibinfo {title} {Excitons in boron nitride single layer},}\
  }\href@noop {} {\bibfield  {journal} {\bibinfo  {journal} {Phys. Rev. B}\
  }\textbf {\bibinfo {volume} {94}},\ \bibinfo {pages} {125303} (\bibinfo
  {year} {2016})}\BibitemShut {NoStop}%
\bibitem [{\citenamefont {Glazov}\ \emph {et~al.}(2017)\citenamefont {Glazov},
  \citenamefont {Golub}, \citenamefont {Wang}, \citenamefont {Marie},
  \citenamefont {Amand},\ and\ \citenamefont {Urbaszek}}]{Glazov2017}%
  \BibitemOpen
  \bibfield  {author} {\bibinfo {author} {\bibfnamefont {M.~M.}\ \bibnamefont
  {Glazov}}, \bibinfo {author} {\bibfnamefont {L.~E.}\ \bibnamefont {Golub}},
  \bibinfo {author} {\bibfnamefont {G.}~\bibnamefont {Wang}}, \bibinfo {author}
  {\bibfnamefont {X.}~\bibnamefont {Marie}}, \bibinfo {author} {\bibfnamefont
  {T.}~\bibnamefont {Amand}}, \ and\ \bibinfo {author} {\bibfnamefont
  {B.}~\bibnamefont {Urbaszek}},\ }\bibfield  {title} {\enquote {\bibinfo
  {title} {Intrinsic exciton-state mixing and nonlinear optical properties in
  transition metal dichalcogenide monolayers},}\ }\href {\doibase
  10.1103/PhysRevB.95.035311} {\bibfield  {journal} {\bibinfo  {journal} {Phys.
  Rev. B}\ }\textbf {\bibinfo {volume} {95}},\ \bibinfo {pages} {035311}
  (\bibinfo {year} {2017})}\BibitemShut {NoStop}%
\bibitem [{\citenamefont {Gong}\ \emph {et~al.}(2017)\citenamefont {Gong},
  \citenamefont {Yu}, \citenamefont {Wang},\ and\ \citenamefont
  {Yao}}]{Gong2017}%
  \BibitemOpen
  \bibfield  {author} {\bibinfo {author} {\bibfnamefont {P.}~\bibnamefont
  {Gong}}, \bibinfo {author} {\bibfnamefont {H.}~\bibnamefont {Yu}}, \bibinfo
  {author} {\bibfnamefont {Y.}~\bibnamefont {Wang}}, \ and\ \bibinfo {author}
  {\bibfnamefont {W.}~\bibnamefont {Yao}},\ }\bibfield  {title} {\enquote
  {\bibinfo {title} {Optical selection rules for excitonic rydberg series in
  the massive dirac cones of hexagonal two-dimensional materials},}\ }\href
  {\doibase 10.1103/PhysRevB.95.125420} {\bibfield  {journal} {\bibinfo
  {journal} {Phys. Rev. B}\ }\textbf {\bibinfo {volume} {95}},\ \bibinfo
  {pages} {125420} (\bibinfo {year} {2017})}\BibitemShut {NoStop}%
\bibitem [{\citenamefont {Attaccalite}\ and\ \citenamefont
  {Gr{\"u}ning}(2013)}]{Attaccalite2013}%
  \BibitemOpen
  \bibfield  {author} {\bibinfo {author} {\bibfnamefont {C}~\bibnamefont
  {Attaccalite}}\ and\ \bibinfo {author} {\bibfnamefont {M}~\bibnamefont
  {Gr{\"u}ning}},\ }\bibfield  {title} {\enquote {\bibinfo {title} {Nonlinear
  optics from an ab initio approach by means of the dynamical berry phase:
  Application to second-and third-harmonic generation in semiconductors},}\
  }\href@noop {} {\bibfield  {journal} {\bibinfo  {journal} {Phys. Rev. B}\
  }\textbf {\bibinfo {volume} {88}},\ \bibinfo {pages} {235113} (\bibinfo
  {year} {2013})}\BibitemShut {NoStop}%
\bibitem [{\citenamefont {Watanabe}\ and\ \citenamefont
  {Taniguchi}(2009)}]{Watanabe2009}%
  \BibitemOpen
  \bibfield  {author} {\bibinfo {author} {\bibfnamefont {K.}~\bibnamefont
  {Watanabe}}\ and\ \bibinfo {author} {\bibfnamefont {T.}~\bibnamefont
  {Taniguchi}},\ }\bibfield  {title} {\enquote {\bibinfo {title} {{Jahn-Teller
  effect on exciton states in hexagonal boron nitride single crystal}},}\
  }\href@noop {} {\bibfield  {journal} {\bibinfo  {journal} {Phys. Rev. B}\
  }\textbf {\bibinfo {volume} {79}},\ \bibinfo {pages} {193104} (\bibinfo
  {year} {2009})}\BibitemShut {NoStop}%
\bibitem [{\citenamefont {Watanabe}\ and\ \citenamefont
  {Taniguchi}(2011)}]{Watanabe2011}%
  \BibitemOpen
  \bibfield  {author} {\bibinfo {author} {\bibfnamefont {K.}~\bibnamefont
  {Watanabe}}\ and\ \bibinfo {author} {\bibfnamefont {T.}~\bibnamefont
  {Taniguchi}},\ }\bibfield  {title} {\enquote {\bibinfo {title} {Hexagonal
  boron nitride as a new ultraviolet luminescent material and its
  application},}\ }\href {\doibase 10.1111/j.1744-7402.2011.02626.x} {\bibfield
   {journal} {\bibinfo  {journal} {International Journal of Applied Ceramic
  Technology}\ }\textbf {\bibinfo {volume} {8}},\ \bibinfo {pages} {977--989}
  (\bibinfo {year} {2011})}\BibitemShut {NoStop}%
\bibitem [{\citenamefont {Jaffrennou}\ \emph {et~al.}(2007)\citenamefont
  {Jaffrennou}, \citenamefont {Barjon}, \citenamefont {Lauret}, \citenamefont
  {Loiseau}, \citenamefont {Ducastelle},\ and\ \citenamefont
  {Attal-Tretout}}]{Jaffrennou2007}%
  \BibitemOpen
  \bibfield  {author} {\bibinfo {author} {\bibfnamefont {P.}~\bibnamefont
  {Jaffrennou}}, \bibinfo {author} {\bibfnamefont {J.}~\bibnamefont {Barjon}},
  \bibinfo {author} {\bibfnamefont {J.-S.}\ \bibnamefont {Lauret}}, \bibinfo
  {author} {\bibfnamefont {A.}~\bibnamefont {Loiseau}}, \bibinfo {author}
  {\bibfnamefont {F.}~\bibnamefont {Ducastelle}}, \ and\ \bibinfo {author}
  {\bibfnamefont {B.}~\bibnamefont {Attal-Tretout}},\ }\bibfield  {title}
  {\enquote {\bibinfo {title} {{Origin of the excitonic recombinations in
  hexagonal boron nitride by spatially resolved cathodoluminescence
  spectroscopy}},}\ }\href@noop {} {\bibfield  {journal} {\bibinfo  {journal}
  {Journal of Applied Physics}\ }\textbf {\bibinfo {volume} {102}},\ \bibinfo
  {pages} {116102} (\bibinfo {year} {2007})}\BibitemShut {NoStop}%
\bibitem [{\citenamefont {Museur}\ \emph {et~al.}(2011)\citenamefont {Museur},
  \citenamefont {Brasse}, \citenamefont {Pierret}, \citenamefont {Maine},
  \citenamefont {Attal-Tretout}, \citenamefont {Ducastelle}, \citenamefont
  {Loiseau}, \citenamefont {Barjon}, \citenamefont {Watanabe}, \citenamefont
  {Taniguchi},\ and\ \citenamefont {Kanaev}}]{Museur2011}%
  \BibitemOpen
  \bibfield  {author} {\bibinfo {author} {\bibfnamefont {L.}~\bibnamefont
  {Museur}}, \bibinfo {author} {\bibfnamefont {G.}~\bibnamefont {Brasse}},
  \bibinfo {author} {\bibfnamefont {A.}~\bibnamefont {Pierret}}, \bibinfo
  {author} {\bibfnamefont {S.}~\bibnamefont {Maine}}, \bibinfo {author}
  {\bibfnamefont {B.}~\bibnamefont {Attal-Tretout}}, \bibinfo {author}
  {\bibfnamefont {F.}~\bibnamefont {Ducastelle}}, \bibinfo {author}
  {\bibfnamefont {A.}~\bibnamefont {Loiseau}}, \bibinfo {author} {\bibfnamefont
  {J.}~\bibnamefont {Barjon}}, \bibinfo {author} {\bibfnamefont
  {K.}~\bibnamefont {Watanabe}}, \bibinfo {author} {\bibfnamefont
  {T.}~\bibnamefont {Taniguchi}}, \ and\ \bibinfo {author} {\bibfnamefont
  {A.}~\bibnamefont {Kanaev}},\ }\bibfield  {title} {\enquote {\bibinfo {title}
  {{Exciton optical transitions in a hexagonal boron nitride single
  crystal}},}\ }\href@noop {} {\bibfield  {journal} {\bibinfo  {journal} {Phys.
  status solidi - Rapid Res. Lett.}\ }\textbf {\bibinfo {volume} {5}},\
  \bibinfo {pages} {214--216} (\bibinfo {year} {2011})}\BibitemShut {NoStop}%
\bibitem [{\citenamefont {Pierret}\ \emph {et~al.}(2014)\citenamefont
  {Pierret}, \citenamefont {Loayza}, \citenamefont {Berini}, \citenamefont
  {Betz}, \citenamefont {Pla\c{c}ais}, \citenamefont {Ducastelle},
  \citenamefont {Barjon},\ and\ \citenamefont {Loiseau}}]{Pierret2014}%
  \BibitemOpen
  \bibfield  {author} {\bibinfo {author} {\bibfnamefont {A.}~\bibnamefont
  {Pierret}}, \bibinfo {author} {\bibfnamefont {J.}~\bibnamefont {Loayza}},
  \bibinfo {author} {\bibfnamefont {B.}~\bibnamefont {Berini}}, \bibinfo
  {author} {\bibfnamefont {A.}~\bibnamefont {Betz}}, \bibinfo {author}
  {\bibfnamefont {B.}~\bibnamefont {Pla\c{c}ais}}, \bibinfo {author}
  {\bibfnamefont {F.}~\bibnamefont {Ducastelle}}, \bibinfo {author}
  {\bibfnamefont {J.}~\bibnamefont {Barjon}}, \ and\ \bibinfo {author}
  {\bibfnamefont {A}~\bibnamefont {Loiseau}},\ }\bibfield  {title} {\enquote
  {\bibinfo {title} {{Excitonic recombinations in hBN : From bulk to exfoliated
  layers}},}\ }\href@noop {} {\bibfield  {journal} {\bibinfo  {journal} {Phys.
  Rev. B}\ }\textbf {\bibinfo {volume} {89}},\ \bibinfo {pages} {035414}
  (\bibinfo {year} {2014})}\BibitemShut {NoStop}%
\bibitem [{\citenamefont {Schue}\ \emph {et~al.}(2016)\citenamefont {Schue},
  \citenamefont {Berini}, \citenamefont {Betz}, \citenamefont {Pla\c{c}ais},
  \citenamefont {Ducastelle}, \citenamefont {Barjon},\ and\ \citenamefont
  {Loiseau}}]{Schue2016}%
  \BibitemOpen
  \bibfield  {author} {\bibinfo {author} {\bibfnamefont {L.}~\bibnamefont
  {Schue}}, \bibinfo {author} {\bibfnamefont {B.}~\bibnamefont {Berini}},
  \bibinfo {author} {\bibfnamefont {A.~C.}\ \bibnamefont {Betz}}, \bibinfo
  {author} {\bibfnamefont {B.}~\bibnamefont {Pla\c{c}ais}}, \bibinfo {author}
  {\bibfnamefont {F.}~\bibnamefont {Ducastelle}}, \bibinfo {author}
  {\bibfnamefont {J.}~\bibnamefont {Barjon}}, \ and\ \bibinfo {author}
  {\bibfnamefont {A.}~\bibnamefont {Loiseau}},\ }\bibfield  {title} {\enquote
  {\bibinfo {title} {Dimensionality effects on the luminescence properties of
  hbn},}\ }\href {\doibase 10.1039/C6NR01253A} {\bibfield  {journal} {\bibinfo
  {journal} {Nanoscale}\ }\textbf {\bibinfo {volume} {8}},\ \bibinfo {pages}
  {6986--6993} (\bibinfo {year} {2016})}\BibitemShut {NoStop}%
\bibitem [{\citenamefont {Li}\ \emph {et~al.}(2016)\citenamefont {Li},
  \citenamefont {Cao}, \citenamefont {Hoffman}, \citenamefont {Edgar},
  \citenamefont {Lin},\ and\ \citenamefont {Jiang}}]{Li2016}%
  \BibitemOpen
  \bibfield  {author} {\bibinfo {author} {\bibfnamefont {J.}~\bibnamefont
  {Li}}, \bibinfo {author} {\bibfnamefont {X.~K.}\ \bibnamefont {Cao}},
  \bibinfo {author} {\bibfnamefont {T.~B.}\ \bibnamefont {Hoffman}}, \bibinfo
  {author} {\bibfnamefont {J.~H.}\ \bibnamefont {Edgar}}, \bibinfo {author}
  {\bibfnamefont {J.~Y.}\ \bibnamefont {Lin}}, \ and\ \bibinfo {author}
  {\bibfnamefont {H.~X.}\ \bibnamefont {Jiang}},\ }\bibfield  {title} {\enquote
  {\bibinfo {title} {Nature of exciton transitions in hexagonal boron
  nitride},}\ }\href@noop {} {\bibfield  {journal} {\bibinfo  {journal}
  {Applied Physics Letters}\ }\textbf {\bibinfo {volume} {108}},\ \bibinfo
  {pages} {122101} (\bibinfo {year} {2016})}\BibitemShut {NoStop}%
\bibitem [{\citenamefont {Vuong}\ \emph {et~al.}(2017)\citenamefont {Vuong},
  \citenamefont {Cassabois}, \citenamefont {Valvin}, \citenamefont {Jacques},
  \citenamefont {Cusc\'o}, \citenamefont {Art\'us},\ and\ \citenamefont
  {Gil}}]{Vuong2017}%
  \BibitemOpen
  \bibfield  {author} {\bibinfo {author} {\bibfnamefont {T.~Q.~P.}\
  \bibnamefont {Vuong}}, \bibinfo {author} {\bibfnamefont {G.}~\bibnamefont
  {Cassabois}}, \bibinfo {author} {\bibfnamefont {P.}~\bibnamefont {Valvin}},
  \bibinfo {author} {\bibfnamefont {V.}~\bibnamefont {Jacques}}, \bibinfo
  {author} {\bibfnamefont {R.}~\bibnamefont {Cusc\'o}}, \bibinfo {author}
  {\bibfnamefont {L.}~\bibnamefont {Art\'us}}, \ and\ \bibinfo {author}
  {\bibfnamefont {B.}~\bibnamefont {Gil}},\ }\bibfield  {title} {\enquote
  {\bibinfo {title} {Overtones of interlayer shear modes in the phonon-assisted
  emission spectrum of hexagonal boron nitride},}\ }\href {\doibase
  10.1103/PhysRevB.95.045207} {\bibfield  {journal} {\bibinfo  {journal} {Phys.
  Rev. B}\ }\textbf {\bibinfo {volume} {95}},\ \bibinfo {pages} {045207}
  (\bibinfo {year} {2017})}\BibitemShut {NoStop}%
\bibitem [{\citenamefont {Galambosi}\ \emph {et~al.}(2011)\citenamefont
  {Galambosi}, \citenamefont {Wirtz}, \citenamefont {Soininen}, \citenamefont
  {Serrano}, \citenamefont {Marini}, \citenamefont {Watanabe}, \citenamefont
  {Taniguchi}, \citenamefont {Huotari}, \citenamefont {Rubio},\ and\
  \citenamefont {H\"am\"al\"ainen}}]{Galombosi2011}%
  \BibitemOpen
  \bibfield  {author} {\bibinfo {author} {\bibfnamefont {S.}~\bibnamefont
  {Galambosi}}, \bibinfo {author} {\bibfnamefont {L.}~\bibnamefont {Wirtz}},
  \bibinfo {author} {\bibfnamefont {J.~A.}\ \bibnamefont {Soininen}}, \bibinfo
  {author} {\bibfnamefont {J.}~\bibnamefont {Serrano}}, \bibinfo {author}
  {\bibfnamefont {A.}~\bibnamefont {Marini}}, \bibinfo {author} {\bibfnamefont
  {K.}~\bibnamefont {Watanabe}}, \bibinfo {author} {\bibfnamefont
  {T.}~\bibnamefont {Taniguchi}}, \bibinfo {author} {\bibfnamefont
  {S.}~\bibnamefont {Huotari}}, \bibinfo {author} {\bibfnamefont
  {A.}~\bibnamefont {Rubio}}, \ and\ \bibinfo {author} {\bibfnamefont
  {K.}~\bibnamefont {H\"am\"al\"ainen}},\ }\bibfield  {title} {\enquote
  {\bibinfo {title} {Anisotropic excitonic effects in the energy loss function
  of hexagonal boron nitride},}\ }\href {\doibase 10.1103/PhysRevB.83.081413}
  {\bibfield  {journal} {\bibinfo  {journal} {Phys. Rev. B}\ }\textbf {\bibinfo
  {volume} {83}},\ \bibinfo {pages} {081413} (\bibinfo {year}
  {2011})}\BibitemShut {NoStop}%
\bibitem [{\citenamefont {Fugallo}\ \emph {et~al.}(2015)\citenamefont
  {Fugallo}, \citenamefont {Aramini}, \citenamefont {Koskelo}, \citenamefont
  {Watanabe}, \citenamefont {Taniguchi}, \citenamefont {Hakala}, \citenamefont
  {Huotari}, \citenamefont {Gatti},\ and\ \citenamefont
  {Sottile}}]{Fugallo2015}%
  \BibitemOpen
  \bibfield  {author} {\bibinfo {author} {\bibfnamefont {G.}~\bibnamefont
  {Fugallo}}, \bibinfo {author} {\bibfnamefont {M.}~\bibnamefont {Aramini}},
  \bibinfo {author} {\bibfnamefont {J.}~\bibnamefont {Koskelo}}, \bibinfo
  {author} {\bibfnamefont {K.}~\bibnamefont {Watanabe}}, \bibinfo {author}
  {\bibfnamefont {T.}~\bibnamefont {Taniguchi}}, \bibinfo {author}
  {\bibfnamefont {M.}~\bibnamefont {Hakala}}, \bibinfo {author} {\bibfnamefont
  {S.}~\bibnamefont {Huotari}}, \bibinfo {author} {\bibfnamefont
  {M.}~\bibnamefont {Gatti}}, \ and\ \bibinfo {author} {\bibfnamefont
  {F.}~\bibnamefont {Sottile}},\ }\bibfield  {title} {\enquote {\bibinfo
  {title} {Exciton energy-momentum map of hexagonal boron nitride},}\ }\href
  {\doibase 10.1103/PhysRevB.92.165122} {\bibfield  {journal} {\bibinfo
  {journal} {Phys. Rev. B}\ }\textbf {\bibinfo {volume} {92}},\ \bibinfo
  {pages} {165122} (\bibinfo {year} {2015})}\BibitemShut {NoStop}%
\bibitem [{\citenamefont {Fossard}\ \emph {et~al.}(2017)\citenamefont
  {Fossard}, \citenamefont {Sponza}, \citenamefont {Schu\'e}, \citenamefont
  {Attaccalite}, \citenamefont {Ducastelle}, \citenamefont {Barjon},\ and\
  \citenamefont {Loiseau}}]{Fossard2017}%
  \BibitemOpen
  \bibfield  {author} {\bibinfo {author} {\bibfnamefont {F.}~\bibnamefont
  {Fossard}}, \bibinfo {author} {\bibfnamefont {L.}~\bibnamefont {Sponza}},
  \bibinfo {author} {\bibfnamefont {L.}~\bibnamefont {Schu\'e}}, \bibinfo
  {author} {\bibfnamefont {C.}~\bibnamefont {Attaccalite}}, \bibinfo {author}
  {\bibfnamefont {F.}~\bibnamefont {Ducastelle}}, \bibinfo {author}
  {\bibfnamefont {J.}~\bibnamefont {Barjon}}, \ and\ \bibinfo {author}
  {\bibfnamefont {A.}~\bibnamefont {Loiseau}},\ }\bibfield  {title} {\enquote
  {\bibinfo {title} {Angle-resolved electron energy loss spectroscopy in
  hexagonal boron nitride},}\ }\href {\doibase 10.1103/PhysRevB.96.115304}
  {\bibfield  {journal} {\bibinfo  {journal} {Phys. Rev. B}\ }\textbf {\bibinfo
  {volume} {96}},\ \bibinfo {pages} {115304} (\bibinfo {year}
  {2017})}\BibitemShut {NoStop}%
\bibitem [{\citenamefont {Schuster}\ \emph {et~al.}(2018)\citenamefont
  {Schuster}, \citenamefont {Habenicht}, \citenamefont {Ahmad}, \citenamefont
  {Knupfer},\ and\ \citenamefont {B{\"u}chner}}]{Schuster2017}%
  \BibitemOpen
  \bibfield  {author} {\bibinfo {author} {\bibfnamefont {R}~\bibnamefont
  {Schuster}}, \bibinfo {author} {\bibfnamefont {C}~\bibnamefont {Habenicht}},
  \bibinfo {author} {\bibfnamefont {M}~\bibnamefont {Ahmad}}, \bibinfo {author}
  {\bibfnamefont {M}~\bibnamefont {Knupfer}}, \ and\ \bibinfo {author}
  {\bibfnamefont {B}~\bibnamefont {B{\"u}chner}},\ }\bibfield  {title}
  {\enquote {\bibinfo {title} {Direct observation of the lowest indirect
  exciton state in the bulk of hexagonal boron nitride},}\ }\href@noop {}
  {\bibfield  {journal} {\bibinfo  {journal} {Physical Review B}\ }\textbf
  {\bibinfo {volume} {97}},\ \bibinfo {pages} {041201} (\bibinfo {year}
  {2018})}\BibitemShut {NoStop}%
\bibitem [{\citenamefont {Sponza}\ \emph {et~al.}(2018)\citenamefont {Sponza},
  \citenamefont {Amara}, \citenamefont {Ducastelle}, \citenamefont {Loiseau},\
  and\ \citenamefont {Attaccalite}}]{Sponza2017}%
  \BibitemOpen
  \bibfield  {author} {\bibinfo {author} {\bibfnamefont {Lorenzo}\ \bibnamefont
  {Sponza}}, \bibinfo {author} {\bibfnamefont {Hakim}\ \bibnamefont {Amara}},
  \bibinfo {author} {\bibfnamefont {Fran{\c{c}}ois}\ \bibnamefont
  {Ducastelle}}, \bibinfo {author} {\bibfnamefont {Annick}\ \bibnamefont
  {Loiseau}}, \ and\ \bibinfo {author} {\bibfnamefont {Claudio}\ \bibnamefont
  {Attaccalite}},\ }\bibfield  {title} {\enquote {\bibinfo {title} {Exciton
  interference in hexagonal boron nitride},}\ }\href@noop {} {\bibfield
  {journal} {\bibinfo  {journal} {Physical Review B}\ }\textbf {\bibinfo
  {volume} {97}},\ \bibinfo {pages} {075121} (\bibinfo {year}
  {2018})}\BibitemShut {NoStop}%
\bibitem [{\citenamefont {Arnaud}\ \emph {et~al.}(2006)\citenamefont {Arnaud},
  \citenamefont {Leb\`{e}gue}, \citenamefont {Rabiller},\ and\ \citenamefont
  {Alouani}}]{Arnaud2006}%
  \BibitemOpen
  \bibfield  {author} {\bibinfo {author} {\bibfnamefont {B.}~\bibnamefont
  {Arnaud}}, \bibinfo {author} {\bibfnamefont {S.}~\bibnamefont {Leb\`{e}gue}},
  \bibinfo {author} {\bibfnamefont {P.}~\bibnamefont {Rabiller}}, \ and\
  \bibinfo {author} {\bibfnamefont {M.}~\bibnamefont {Alouani}},\ }\bibfield
  {title} {\enquote {\bibinfo {title} {{Huge excitonic effects in layered
  hexagonal boron nitride}},}\ }\href@noop {} {\bibfield  {journal} {\bibinfo
  {journal} {Physical Review Letters}\ }\textbf {\bibinfo {volume} {96}},\
  \bibinfo {pages} {026402} (\bibinfo {year} {2006})}\BibitemShut {NoStop}%
\bibitem [{\citenamefont {Arnaud}\ \emph {et~al.}(2008)\citenamefont {Arnaud},
  \citenamefont {Leb\`egue}, \citenamefont {Rabiller},\ and\ \citenamefont
  {Alouani}}]{Arnaud2008}%
  \BibitemOpen
  \bibfield  {author} {\bibinfo {author} {\bibfnamefont {B.}~\bibnamefont
  {Arnaud}}, \bibinfo {author} {\bibfnamefont {S.}~\bibnamefont {Leb\`egue}},
  \bibinfo {author} {\bibfnamefont {P.}~\bibnamefont {Rabiller}}, \ and\
  \bibinfo {author} {\bibfnamefont {M.}~\bibnamefont {Alouani}},\ }\bibfield
  {title} {\enquote {\bibinfo {title} {Arnaud, leb\`egue, rabiller, and alouani
  reply:},}\ }\href {\doibase 10.1103/PhysRevLett.100.189702} {\bibfield
  {journal} {\bibinfo  {journal} {Phys. Rev. Lett.}\ }\textbf {\bibinfo
  {volume} {100}},\ \bibinfo {pages} {189702} (\bibinfo {year}
  {2008})}\BibitemShut {NoStop}%
\bibitem [{\citenamefont {Wirtz}\ \emph {et~al.}(2006)\citenamefont {Wirtz},
  \citenamefont {Marini},\ and\ \citenamefont {Rubio}}]{Wirtz2006}%
  \BibitemOpen
  \bibfield  {author} {\bibinfo {author} {\bibfnamefont {L.}~\bibnamefont
  {Wirtz}}, \bibinfo {author} {\bibfnamefont {A.}~\bibnamefont {Marini}}, \
  and\ \bibinfo {author} {\bibfnamefont {A.}~\bibnamefont {Rubio}},\ }\bibfield
   {title} {\enquote {\bibinfo {title} {Excitons in boron nitride nanotubes:
  dimensionality effects},}\ }\href@noop {} {\bibfield  {journal} {\bibinfo
  {journal} {Phys. Rev. Lett.}\ }\textbf {\bibinfo {volume} {96}},\ \bibinfo
  {pages} {126104} (\bibinfo {year} {2006})}\BibitemShut {NoStop}%
\bibitem [{\citenamefont {Wirtz}\ \emph {et~al.}(2008)\citenamefont {Wirtz},
  \citenamefont {Marini}, \citenamefont {Gr\"uning}, \citenamefont
  {Attaccalite}, \citenamefont {Kresse},\ and\ \citenamefont
  {Rubio}}]{Wirtz2008}%
  \BibitemOpen
  \bibfield  {author} {\bibinfo {author} {\bibfnamefont {L.}~\bibnamefont
  {Wirtz}}, \bibinfo {author} {\bibfnamefont {A.}~\bibnamefont {Marini}},
  \bibinfo {author} {\bibfnamefont {M.}~\bibnamefont {Gr\"uning}}, \bibinfo
  {author} {\bibfnamefont {C.}~\bibnamefont {Attaccalite}}, \bibinfo {author}
  {\bibfnamefont {G.}~\bibnamefont {Kresse}}, \ and\ \bibinfo {author}
  {\bibfnamefont {A.}~\bibnamefont {Rubio}},\ }\bibfield  {title} {\enquote
  {\bibinfo {title} {Comment on ``huge excitonic effects in layered hexagonal
  boron nitride''},}\ }\href {\doibase 10.1103/PhysRevLett.100.189701}
  {\bibfield  {journal} {\bibinfo  {journal} {Phys. Rev. Lett.}\ }\textbf
  {\bibinfo {volume} {100}},\ \bibinfo {pages} {189701} (\bibinfo {year}
  {2008})}\BibitemShut {NoStop}%
\bibitem [{\citenamefont {Schue}\ \emph {et~al.}(2018)\citenamefont {Schue},
  \citenamefont {Sponza}, \citenamefont {Plaud}, \citenamefont {Bensalah},
  \citenamefont {Watanabe}, \citenamefont {Taniguchi}, \citenamefont
  {Ducastelle}, \citenamefont {Loiseau},\ and\ \citenamefont
  {Barjon}}]{Schue2018}%
  \BibitemOpen
  \bibfield  {author} {\bibinfo {author} {\bibfnamefont {Leonard}\ \bibnamefont
  {Schue}}, \bibinfo {author} {\bibfnamefont {Lorenzo}\ \bibnamefont {Sponza}},
  \bibinfo {author} {\bibfnamefont {Alexandre}\ \bibnamefont {Plaud}}, \bibinfo
  {author} {\bibfnamefont {Hakima}\ \bibnamefont {Bensalah}}, \bibinfo {author}
  {\bibfnamefont {Kenji}\ \bibnamefont {Watanabe}}, \bibinfo {author}
  {\bibfnamefont {Takashi}\ \bibnamefont {Taniguchi}}, \bibinfo {author}
  {\bibfnamefont {Fran{\c c}ois}\ \bibnamefont {Ducastelle}}, \bibinfo {author}
  {\bibfnamefont {Annick}\ \bibnamefont {Loiseau}}, \ and\ \bibinfo {author}
  {\bibfnamefont {Julien}\ \bibnamefont {Barjon}},\ }\bibfield  {title}
  {\enquote {\bibinfo {title} {Direct and indirect excitons with high binding
  energies in hbn},}\ }\href@noop {} {\bibfield  {journal} {\bibinfo  {journal}
  {arXiv:1803.03766 [cond-mat.mtrl-sci]}\ } (\bibinfo {year}
  {2018})}\BibitemShut {NoStop}%
\bibitem [{\citenamefont {Paleari}\ \emph {et~al.}(2018)\citenamefont
  {Paleari}, \citenamefont {Galvani}, \citenamefont {Amara}, \citenamefont
  {Ducastelle}, \citenamefont {Molina-S{\'a}nchez},\ and\ \citenamefont
  {Wirtz}}]{Paleari2018}%
  \BibitemOpen
  \bibfield  {author} {\bibinfo {author} {\bibfnamefont {F.}~\bibnamefont
  {Paleari}}, \bibinfo {author} {\bibfnamefont {T.}~\bibnamefont {Galvani}},
  \bibinfo {author} {\bibfnamefont {H.}~\bibnamefont {Amara}}, \bibinfo
  {author} {\bibfnamefont {F.}~\bibnamefont {Ducastelle}}, \bibinfo {author}
  {\bibfnamefont {A.}~\bibnamefont {Molina-S{\'a}nchez}}, \ and\ \bibinfo
  {author} {\bibfnamefont {L.}~\bibnamefont {Wirtz}},\ }\bibfield  {title}
  {\enquote {\bibinfo {title} {Excitons in few-layer hexagonal boron nitride:
  Davydov splitting and surface localization},}\ }\href@noop {} {\bibfield
  {journal} {\bibinfo  {journal} {arXiv:1803.00982 [cond-mat.mtrl-sci]}\ }
  (\bibinfo {year} {2018})}\BibitemShut {NoStop}%
\bibitem [{\citenamefont {Blase}\ \emph {et~al.}(1995)\citenamefont {Blase},
  \citenamefont {Rubio}, \citenamefont {Louie},\ and\ \citenamefont
  {Cohen}}]{Blase1995}%
  \BibitemOpen
  \bibfield  {author} {\bibinfo {author} {\bibfnamefont {X.}~\bibnamefont
  {Blase}}, \bibinfo {author} {\bibfnamefont {Angel}\ \bibnamefont {Rubio}},
  \bibinfo {author} {\bibfnamefont {Steven~G.}\ \bibnamefont {Louie}}, \ and\
  \bibinfo {author} {\bibfnamefont {Marvin~L.}\ \bibnamefont {Cohen}},\
  }\bibfield  {title} {\enquote {\bibinfo {title} {Quasiparticle band structure
  of bulk hexagonal boron nitride and related systems},}\ }\href {\doibase
  10.1103/PhysRevB.51.6868} {\bibfield  {journal} {\bibinfo  {journal} {Phys.
  Rev. B}\ }\textbf {\bibinfo {volume} {51}},\ \bibinfo {pages} {6868--6875}
  (\bibinfo {year} {1995})}\BibitemShut {NoStop}%
\bibitem [{\citenamefont {Solozhenko}\ \emph {et~al.}(1995)\citenamefont
  {Solozhenko}, \citenamefont {Will},\ and\ \citenamefont
  {Elf}}]{solozhenko1995isothermal}%
  \BibitemOpen
  \bibfield  {author} {\bibinfo {author} {\bibfnamefont {VL}~\bibnamefont
  {Solozhenko}}, \bibinfo {author} {\bibfnamefont {G}~\bibnamefont {Will}}, \
  and\ \bibinfo {author} {\bibfnamefont {F}~\bibnamefont {Elf}},\ }\bibfield
  {title} {\enquote {\bibinfo {title} {Isothermal compression of hexagonal
  graphite-like boron nitride up to 12 gpa},}\ }\href@noop {} {\bibfield
  {journal} {\bibinfo  {journal} {Solid state communications}\ }\textbf
  {\bibinfo {volume} {96}},\ \bibinfo {pages} {1--3} (\bibinfo {year}
  {1995})}\BibitemShut {NoStop}%
\bibitem [{Note1()}]{Note1}%
  \BibitemOpen
  \bibinfo {note} {In the case of dichalcogenides the symmetry analysis is more
  complex because of the presence of $d$ orbitals of different
  symmetries.}\BibitemShut {Stop}%
\bibitem [{Note2()}]{Note2}%
  \BibitemOpen
  \bibinfo {note} {The matrix element of the velocity operator does not depend
  on the Coulomb potential which has been assumed to be local, and therefore
  commutes with the $\protect \textbf {r}$ operator.}\BibitemShut {Stop}%
\bibitem [{\citenamefont {Yu}\ and\ \citenamefont {Cardona}(2010)}]{Yu2010}%
  \BibitemOpen
  \bibfield  {author} {\bibinfo {author} {\bibfnamefont {P.~Y.}\ \bibnamefont
  {Yu}}\ and\ \bibinfo {author} {\bibfnamefont {M.}~\bibnamefont {Cardona}},\
  }\href@noop {} {\emph {\bibinfo {title} {Fundamentals of Semiconductors}}}\
  (\bibinfo  {publisher} {Springer},\ \bibinfo {year} {2010})\BibitemShut
  {NoStop}%
\bibitem [{\citenamefont {Toyozawa}(2003)}]{Toyozawa2003}%
  \BibitemOpen
  \bibfield  {author} {\bibinfo {author} {\bibfnamefont {Y.}~\bibnamefont
  {Toyozawa}},\ }\href@noop {} {\emph {\bibinfo {title} {Optical Processes in
  Solids}}}\ (\bibinfo  {publisher} {Cambridge University Press},\ \bibinfo
  {year} {2003})\BibitemShut {NoStop}%
\bibitem [{\citenamefont {Grosso}\ and\ \citenamefont
  {Parravicini}(2014)}]{Grosso2014}%
  \BibitemOpen
  \bibfield  {author} {\bibinfo {author} {\bibfnamefont {G.}~\bibnamefont
  {Grosso}}\ and\ \bibinfo {author} {\bibfnamefont {G.}~\bibnamefont
  {Parravicini}},\ }\href@noop {} {\emph {\bibinfo {title} {Solid State
  Physics}}},\ \bibinfo {edition} {2nd}\ ed.\ (\bibinfo  {publisher} {Academic
  Press},\ \bibinfo {year} {2014})\BibitemShut {NoStop}%
\bibitem [{\citenamefont {Boyd}(2008)}]{Boyd2008}%
  \BibitemOpen
  \bibfield  {author} {\bibinfo {author} {\bibfnamefont {R.~W.}\ \bibnamefont
  {Boyd}},\ }\href@noop {} {\emph {\bibinfo {title} {Nonlinear Optics}}},\
  \bibinfo {edition} {3rd}\ ed.\ (\bibinfo  {publisher} {Academic Press},\
  \bibinfo {year} {2008})\BibitemShut {NoStop}%
\bibitem [{\citenamefont {Lin}\ \emph {et~al.}(2007)\citenamefont {Lin},
  \citenamefont {Painter},\ and\ \citenamefont {Agrawal}}]{Lin2007}%
  \BibitemOpen
  \bibfield  {author} {\bibinfo {author} {\bibfnamefont {Q}~\bibnamefont
  {Lin}}, \bibinfo {author} {\bibfnamefont {Oskar~J}\ \bibnamefont {Painter}},
  \ and\ \bibinfo {author} {\bibfnamefont {Govind~P}\ \bibnamefont {Agrawal}},\
  }\bibfield  {title} {\enquote {\bibinfo {title} {Nonlinear optical phenomena
  in silicon waveguides: modeling and applications},}\ }\href@noop {}
  {\bibfield  {journal} {\bibinfo  {journal} {Optics Express}\ }\textbf
  {\bibinfo {volume} {15}},\ \bibinfo {pages} {16604--16644} (\bibinfo {year}
  {2007})}\BibitemShut {NoStop}%
\bibitem [{\citenamefont {Souza}\ \emph {et~al.}(2004)\citenamefont {Souza},
  \citenamefont {\'I\~niguez},\ and\ \citenamefont {Vanderbilt}}]{Souza2004}%
  \BibitemOpen
  \bibfield  {author} {\bibinfo {author} {\bibfnamefont {I.}~\bibnamefont
  {Souza}}, \bibinfo {author} {\bibfnamefont {J.}~\bibnamefont {\'I\~niguez}},
  \ and\ \bibinfo {author} {\bibfnamefont {D.}~\bibnamefont {Vanderbilt}},\
  }\bibfield  {title} {\enquote {\bibinfo {title} {Dynamics of berry-phase
  polarization in time-dependent electric fields},}\ }\href {\doibase
  10.1103/PhysRevB.69.085106} {\bibfield  {journal} {\bibinfo  {journal} {Phys.
  Rev. B}\ }\textbf {\bibinfo {volume} {69}},\ \bibinfo {pages} {085106}
  (\bibinfo {year} {2004})}\BibitemShut {NoStop}%
\bibitem [{\citenamefont {Richardson}\ and\ \citenamefont
  {Gaunt}(1927)}]{Richardson1927}%
  \BibitemOpen
  \bibfield  {author} {\bibinfo {author} {\bibfnamefont {L.~F.}\ \bibnamefont
  {Richardson}}\ and\ \bibinfo {author} {\bibfnamefont {J.~A.}\ \bibnamefont
  {Gaunt}},\ }\bibfield  {title} {\enquote {\bibinfo {title} {{VIII. The
  deferred approach to the limit}},}\ }\href {\doibase 10.1098/rsta.1927.0008}
  {\bibfield  {journal} {\bibinfo  {journal} {Philosophical Transactions of the
  Royal Society of London A: Mathematical, Physical and Engineering Sciences}\
  }\textbf {\bibinfo {volume} {226}},\ \bibinfo {pages} {299--361} (\bibinfo
  {year} {1927})}\BibitemShut {NoStop}%
\bibitem [{\citenamefont {Attaccalite}\ \emph {et~al.}(2011)\citenamefont
  {Attaccalite}, \citenamefont {Gr{\"u}ning},\ and\ \citenamefont
  {Marini}}]{attaccalite2011real}%
  \BibitemOpen
  \bibfield  {author} {\bibinfo {author} {\bibfnamefont {Claudio}\ \bibnamefont
  {Attaccalite}}, \bibinfo {author} {\bibfnamefont {M}~\bibnamefont
  {Gr{\"u}ning}}, \ and\ \bibinfo {author} {\bibfnamefont {A}~\bibnamefont
  {Marini}},\ }\bibfield  {title} {\enquote {\bibinfo {title} {Real-time
  approach to the optical properties of solids and nanostructures:
  Time-dependent bethe-salpeter equation},}\ }\href@noop {} {\bibfield
  {journal} {\bibinfo  {journal} {Physical Review B}\ }\textbf {\bibinfo
  {volume} {84}},\ \bibinfo {pages} {245110} (\bibinfo {year}
  {2011})}\BibitemShut {NoStop}%
\bibitem [{Note3()}]{Note3}%
  \BibitemOpen
  \bibinfo {note} {The Kohn-Sham\ system is a fictitious system of independent
  particles in an effective local field such that the electronic density of the
  physical system is reproduced, see W. Kohn and L. J. Sham, Phys. Rev. 140,
  A1133 (1965)}\BibitemShut {NoStop}%
\bibitem [{\citenamefont {Adler}(1962)}]{Adler1962}%
  \BibitemOpen
  \bibfield  {author} {\bibinfo {author} {\bibfnamefont {S.~L.}\ \bibnamefont
  {Adler}},\ }\bibfield  {title} {\enquote {\bibinfo {title} {Quantum theory of
  the dielectric constant in real solids},}\ }\href {\doibase
  10.1103/PhysRev.126.413} {\bibfield  {journal} {\bibinfo  {journal} {Phys.
  Rev.}\ }\textbf {\bibinfo {volume} {126}},\ \bibinfo {pages} {413--420}
  (\bibinfo {year} {1962})}\BibitemShut {NoStop}%
\bibitem [{\citenamefont {{Strinati}}(1988)}]{Strinati1988}%
  \BibitemOpen
  \bibfield  {author} {\bibinfo {author} {\bibfnamefont {G.}~\bibnamefont
  {{Strinati}}},\ }\bibfield  {title} {\enquote {\bibinfo {title} {{Application
  of the Green's functions method to the study of the optical properties of
  semiconductors}},}\ }\href {\doibase 10.1007/BF02725962} {\bibfield
  {journal} {\bibinfo  {journal} {Nuovo Cimento Rivista Serie}\ }\textbf
  {\bibinfo {volume} {11}},\ \bibinfo {pages} {1--86} (\bibinfo {year}
  {1988})}\BibitemShut {NoStop}%
\bibitem [{\citenamefont {Grynberg}\ \emph {et~al.}(1990)\citenamefont
  {Grynberg}, \citenamefont {Aspect},\ and\ \citenamefont
  {Fabre}}]{Grynberg2010}%
  \BibitemOpen
  \bibfield  {author} {\bibinfo {author} {\bibfnamefont {G.}~\bibnamefont
  {Grynberg}}, \bibinfo {author} {\bibfnamefont {A.}~\bibnamefont {Aspect}}, \
  and\ \bibinfo {author} {\bibfnamefont {C.}~\bibnamefont {Fabre}},\
  }\href@noop {} {\emph {\bibinfo {title} {Introduction to Quantum Optics}}}\
  (\bibinfo  {publisher} {Cambridge University Press},\ \bibinfo {year}
  {1990})\BibitemShut {NoStop}%
\bibitem [{\citenamefont {Mahan}(1968)}]{Mahan1968}%
  \BibitemOpen
  \bibfield  {author} {\bibinfo {author} {\bibfnamefont {G.~D.}\ \bibnamefont
  {Mahan}},\ }\bibfield  {title} {\enquote {\bibinfo {title} {Theory of
  two-photon spectroscopy in solids},}\ }\href {\doibase
  10.1103/PhysRev.170.825} {\bibfield  {journal} {\bibinfo  {journal} {Phys.
  Rev.}\ }\textbf {\bibinfo {volume} {170}},\ \bibinfo {pages} {825--838}
  (\bibinfo {year} {1968})}\BibitemShut {NoStop}%
\bibitem [{\citenamefont {Shimizu}(1989)}]{Shimizu1989}%
  \BibitemOpen
  \bibfield  {author} {\bibinfo {author} {\bibfnamefont {A.}~\bibnamefont
  {Shimizu}},\ }\bibfield  {title} {\enquote {\bibinfo {title} {Two-photon
  absorption in quantum-well structures near half the direct band gap},}\
  }\href {\doibase 10.1103/PhysRevB.40.1403} {\bibfield  {journal} {\bibinfo
  {journal} {Phys. Rev. B}\ }\textbf {\bibinfo {volume} {40}},\ \bibinfo
  {pages} {1403--1406} (\bibinfo {year} {1989})}\BibitemShut {NoStop}%
\bibitem [{Note4()}]{Note4}%
  \BibitemOpen
  \bibinfo {note} {M. Glazov, private communication; this is also discussed in
  Ref.[\protect \rev@citealpnum {Glazov2017}] for the case of TMD.}\BibitemShut
  {Stop}%
\bibitem [{\citenamefont {Koskelo}\ \emph {et~al.}(2017)\citenamefont
  {Koskelo}, \citenamefont {Fugallo}, \citenamefont {Hakala}, \citenamefont
  {Gatti}, \citenamefont {Sottile},\ and\ \citenamefont
  {Cudazzo}}]{Koskelo2017}%
  \BibitemOpen
  \bibfield  {author} {\bibinfo {author} {\bibfnamefont {Jaakko}\ \bibnamefont
  {Koskelo}}, \bibinfo {author} {\bibfnamefont {Giorgia}\ \bibnamefont
  {Fugallo}}, \bibinfo {author} {\bibfnamefont {Mikko}\ \bibnamefont {Hakala}},
  \bibinfo {author} {\bibfnamefont {Matteo}\ \bibnamefont {Gatti}}, \bibinfo
  {author} {\bibfnamefont {Francesco}\ \bibnamefont {Sottile}}, \ and\ \bibinfo
  {author} {\bibfnamefont {Pierluigi}\ \bibnamefont {Cudazzo}},\ }\bibfield
  {title} {\enquote {\bibinfo {title} {Excitons in van der waals materials:
  From monolayer to bulk hexagonal boron nitride},}\ }\href {\doibase
  10.1103/PhysRevB.95.035125} {\bibfield  {journal} {\bibinfo  {journal} {Phys.
  Rev. B}\ }\textbf {\bibinfo {volume} {95}},\ \bibinfo {pages} {035125}
  (\bibinfo {year} {2017})}\BibitemShut {NoStop}%
\bibitem [{\citenamefont {Giannozzi}\ \emph {et~al.}(2009)\citenamefont
  {Giannozzi}, \citenamefont {Baroni}, \citenamefont {Bonini}, \citenamefont
  {Calandra}, \citenamefont {Car}, \citenamefont {Cavazzoni}, \citenamefont
  {Ceresoli}, \citenamefont {Chiarotti}, \citenamefont {Cococcioni},
  \citenamefont {Dabo}, \citenamefont {Dal~Corso}, \citenamefont
  {de~Gironcoli}, \citenamefont {Fabris}, \citenamefont {Fratesi},
  \citenamefont {Gebauer}, \citenamefont {Gerstmann}, \citenamefont
  {Gougoussis}, \citenamefont {Kokalj}, \citenamefont {Lazzeri}, \citenamefont
  {Martin-Samos}, \citenamefont {Marzari}, \citenamefont {Mauri}, \citenamefont
  {Mazzarello}, \citenamefont {Paolini}, \citenamefont {Pasquarello},
  \citenamefont {Paulatto}, \citenamefont {Sbraccia}, \citenamefont {Scandolo},
  \citenamefont {Sclauzero}, \citenamefont {Seitsonen}, \citenamefont
  {Smogunov}, \citenamefont {Umari},\ and\ \citenamefont
  {Wentzcovitch}}]{Quantum_espresso2009}%
  \BibitemOpen
  \bibfield  {author} {\bibinfo {author} {\bibfnamefont {P.}~\bibnamefont
  {Giannozzi}}, \bibinfo {author} {\bibfnamefont {S.}~\bibnamefont {Baroni}},
  \bibinfo {author} {\bibfnamefont {N.}~\bibnamefont {Bonini}}, \bibinfo
  {author} {\bibfnamefont {M.}~\bibnamefont {Calandra}}, \bibinfo {author}
  {\bibfnamefont {R.}~\bibnamefont {Car}}, \bibinfo {author} {\bibfnamefont
  {C.}~\bibnamefont {Cavazzoni}}, \bibinfo {author} {\bibfnamefont
  {D.}~\bibnamefont {Ceresoli}}, \bibinfo {author} {\bibfnamefont {G.~L.}\
  \bibnamefont {Chiarotti}}, \bibinfo {author} {\bibfnamefont {M.}~\bibnamefont
  {Cococcioni}}, \bibinfo {author} {\bibfnamefont {I}~\bibnamefont {Dabo}},
  \bibinfo {author} {\bibfnamefont {A}~\bibnamefont {Dal~Corso}}, \bibinfo
  {author} {\bibfnamefont {S.}~\bibnamefont {de~Gironcoli}}, \bibinfo {author}
  {\bibfnamefont {S.}~\bibnamefont {Fabris}}, \bibinfo {author} {\bibfnamefont
  {G.}~\bibnamefont {Fratesi}}, \bibinfo {author} {\bibfnamefont
  {R.}~\bibnamefont {Gebauer}}, \bibinfo {author} {\bibfnamefont
  {U.}~\bibnamefont {Gerstmann}}, \bibinfo {author} {\bibfnamefont
  {C.}~\bibnamefont {Gougoussis}}, \bibinfo {author} {\bibfnamefont
  {A.}~\bibnamefont {Kokalj}}, \bibinfo {author} {\bibfnamefont
  {M.}~\bibnamefont {Lazzeri}}, \bibinfo {author} {\bibfnamefont
  {L.}~\bibnamefont {Martin-Samos}}, \bibinfo {author} {\bibfnamefont
  {N.}~\bibnamefont {Marzari}}, \bibinfo {author} {\bibfnamefont
  {F.}~\bibnamefont {Mauri}}, \bibinfo {author} {\bibfnamefont
  {M.}~\bibnamefont {Mazzarello}}, \bibinfo {author} {\bibfnamefont
  {S.}~\bibnamefont {Paolini}}, \bibinfo {author} {\bibfnamefont
  {A.}~\bibnamefont {Pasquarello}}, \bibinfo {author} {\bibfnamefont
  {L.}~\bibnamefont {Paulatto}}, \bibinfo {author} {\bibfnamefont
  {C.}~\bibnamefont {Sbraccia}}, \bibinfo {author} {\bibfnamefont
  {S.}~\bibnamefont {Scandolo}}, \bibinfo {author} {\bibfnamefont
  {G.}~\bibnamefont {Sclauzero}}, \bibinfo {author} {\bibfnamefont {A.~P.}\
  \bibnamefont {Seitsonen}}, \bibinfo {author} {\bibfnamefont {A.}~\bibnamefont
  {Smogunov}}, \bibinfo {author} {\bibfnamefont {P.}~\bibnamefont {Umari}}, \
  and\ \bibinfo {author} {\bibfnamefont {R.~M.}\ \bibnamefont {Wentzcovitch}},\
  }\bibfield  {title} {\enquote {\bibinfo {title} {Quantum espresso: a modular
  and open-source software project for quantum simulations of materials},}\
  }\href {http://stacks.iop.org/0953-8984/21/i=39/a=395502} {\bibfield
  {journal} {\bibinfo  {journal} {Journal of Physics: Condensed Matter}\
  }\textbf {\bibinfo {volume} {21}},\ \bibinfo {pages} {395502} (\bibinfo
  {year} {2009})}\BibitemShut {NoStop}%
\bibitem [{Note5()}]{Note5}%
  \BibitemOpen
  \bibinfo {note} {The approach is implemented in the \protect \textsc {Lumen}
  extension of the \protect \textsc {Yambo} many-body code.~\cite {YAMBO2009}
  The source code is available from \protect \texttt
  {http://www.attaccalite.com/lumen/}}\BibitemShut {NoStop}%
\bibitem [{\citenamefont {Koonin}\ and\ \citenamefont
  {Meredith}(1990)}]{Koonin1990}%
  \BibitemOpen
  \bibfield  {author} {\bibinfo {author} {\bibfnamefont {S.~E.}\ \bibnamefont
  {Koonin}}\ and\ \bibinfo {author} {\bibfnamefont {D.~C.}\ \bibnamefont
  {Meredith}},\ }\href@noop {} {\emph {\bibinfo {title} {Computational
  Physics}}}\ (\bibinfo  {publisher} {Addison-Wesley},\ \bibinfo {address}
  {Reading, MA},\ \bibinfo {year} {1990})\BibitemShut {NoStop}%
\bibitem [{\citenamefont {Marini}\ \emph {et~al.}(2009)\citenamefont {Marini},
  \citenamefont {Hogan}, \citenamefont {Gr{\"u}ning},\ and\ \citenamefont
  {Varsano}}]{YAMBO2009}%
  \BibitemOpen
  \bibfield  {author} {\bibinfo {author} {\bibfnamefont {Andrea}\ \bibnamefont
  {Marini}}, \bibinfo {author} {\bibfnamefont {Conor}\ \bibnamefont {Hogan}},
  \bibinfo {author} {\bibfnamefont {Myrta}\ \bibnamefont {Gr{\"u}ning}}, \ and\
  \bibinfo {author} {\bibfnamefont {Daniele}\ \bibnamefont {Varsano}},\
  }\bibfield  {title} {\enquote {\bibinfo {title} {yambo: An ab initio tool for
  excited state calculations},}\ }\href@noop {} {\bibfield  {journal} {\bibinfo
   {journal} {Computer Physics Communications}\ }\textbf {\bibinfo {volume}
  {180}},\ \bibinfo {pages} {1392--1403} (\bibinfo {year} {2009})}\BibitemShut
  {NoStop}%
\bibitem [{Note6()}]{Note6}%
  \BibitemOpen
  \bibinfo {note} {Excitons at higher energy are more difficult to characterize
  because they require a denser $k$-point sampling. As specified above the
  $k$-point samplings are chosen to guarantee the convergence of the first peak
  in the spectra that are the focus of this study.}\BibitemShut {Stop}%
\bibitem [{\citenamefont {S{\"a}yn{\"a}tjoki}\ \emph
  {et~al.}(2017)\citenamefont {S{\"a}yn{\"a}tjoki}, \citenamefont {Karvonen},
  \citenamefont {Rostami}, \citenamefont {Autere}, \citenamefont {Mehravar},
  \citenamefont {Lombardo}, \citenamefont {Norwood}, \citenamefont {Hasan},
  \citenamefont {Peyghambarian}, \citenamefont {Lipsanen} \emph
  {et~al.}}]{saynatjoki2017ultra}%
  \BibitemOpen
  \bibfield  {author} {\bibinfo {author} {\bibfnamefont {Antti}\ \bibnamefont
  {S{\"a}yn{\"a}tjoki}}, \bibinfo {author} {\bibfnamefont {Lasse}\ \bibnamefont
  {Karvonen}}, \bibinfo {author} {\bibfnamefont {Habib}\ \bibnamefont
  {Rostami}}, \bibinfo {author} {\bibfnamefont {Anton}\ \bibnamefont {Autere}},
  \bibinfo {author} {\bibfnamefont {Soroush}\ \bibnamefont {Mehravar}},
  \bibinfo {author} {\bibfnamefont {Antonio}\ \bibnamefont {Lombardo}},
  \bibinfo {author} {\bibfnamefont {Robert~A}\ \bibnamefont {Norwood}},
  \bibinfo {author} {\bibfnamefont {Tawfique}\ \bibnamefont {Hasan}}, \bibinfo
  {author} {\bibfnamefont {Nasser}\ \bibnamefont {Peyghambarian}}, \bibinfo
  {author} {\bibfnamefont {Harri}\ \bibnamefont {Lipsanen}},  \emph {et~al.},\
  }\bibfield  {title} {\enquote {\bibinfo {title} {Ultra-strong nonlinear
  optical processes and trigonal warping in mos 2 layers},}\ }\href@noop {}
  {\bibfield  {journal} {\bibinfo  {journal} {Nature communications}\ }\textbf
  {\bibinfo {volume} {8}},\ \bibinfo {pages} {893} (\bibinfo {year}
  {2017})}\BibitemShut {NoStop}%
\bibitem [{\citenamefont {Zhou}\ \emph {et~al.}(2015)\citenamefont {Zhou},
  \citenamefont {Shan}, \citenamefont {Yao},\ and\ \citenamefont
  {Xiao}}]{Zhou2015}%
  \BibitemOpen
  \bibfield  {author} {\bibinfo {author} {\bibfnamefont {Jianhui}\ \bibnamefont
  {Zhou}}, \bibinfo {author} {\bibfnamefont {Wen-Yu}\ \bibnamefont {Shan}},
  \bibinfo {author} {\bibfnamefont {Wang}\ \bibnamefont {Yao}}, \ and\ \bibinfo
  {author} {\bibfnamefont {Di}~\bibnamefont {Xiao}},\ }\bibfield  {title}
  {\enquote {\bibinfo {title} {Berry phase modification to the energy spectrum
  of excitons},}\ }\href {\doibase 10.1103/PhysRevLett.115.166803} {\bibfield
  {journal} {\bibinfo  {journal} {Phys. Rev. Lett.}\ }\textbf {\bibinfo
  {volume} {115}},\ \bibinfo {pages} {166803} (\bibinfo {year}
  {2015})}\BibitemShut {NoStop}%
\bibitem [{\citenamefont {Srivastava}\ and\ \citenamefont
  {Imamo\u{g}lu}(2015)}]{Srivastava2015}%
  \BibitemOpen
  \bibfield  {author} {\bibinfo {author} {\bibfnamefont {Ajit}\ \bibnamefont
  {Srivastava}}\ and\ \bibinfo {author} {\bibfnamefont {Ata\c{c}}\ \bibnamefont
  {Imamo\u{g}lu}},\ }\bibfield  {title} {\enquote {\bibinfo {title} {Signatures
  of bloch-band geometry on excitons: Nonhydrogenic spectra in transition-metal
  dichalcogenides},}\ }\href {\doibase 10.1103/PhysRevLett.115.166802}
  {\bibfield  {journal} {\bibinfo  {journal} {Phys. Rev. Lett.}\ }\textbf
  {\bibinfo {volume} {115}},\ \bibinfo {pages} {166802} (\bibinfo {year}
  {2015})}\BibitemShut {NoStop}%
\bibitem [{\citenamefont {Cao}\ \emph {et~al.}(2018)\citenamefont {Cao},
  \citenamefont {Wu},\ and\ \citenamefont {Louie}}]{Cao2018}%
  \BibitemOpen
  \bibfield  {author} {\bibinfo {author} {\bibfnamefont {Ting}\ \bibnamefont
  {Cao}}, \bibinfo {author} {\bibfnamefont {Meng}\ \bibnamefont {Wu}}, \ and\
  \bibinfo {author} {\bibfnamefont {Steven~G.}\ \bibnamefont {Louie}},\
  }\bibfield  {title} {\enquote {\bibinfo {title} {Unifying optical selection
  rules for excitons in two dimensions: Band topology and winding numbers},}\
  }\href {\doibase 10.1103/PhysRevLett.120.087402} {\bibfield  {journal}
  {\bibinfo  {journal} {Phys. Rev. Lett.}\ }\textbf {\bibinfo {volume} {120}},\
  \bibinfo {pages} {087402} (\bibinfo {year} {2018})}\BibitemShut {NoStop}%
\bibitem [{\citenamefont {Zhang}\ \emph {et~al.}(2018)\citenamefont {Zhang},
  \citenamefont {Shan},\ and\ \citenamefont {Xiao}}]{Zhang2018}%
  \BibitemOpen
  \bibfield  {author} {\bibinfo {author} {\bibfnamefont {Xiaoou}\ \bibnamefont
  {Zhang}}, \bibinfo {author} {\bibfnamefont {Wen-Yu}\ \bibnamefont {Shan}}, \
  and\ \bibinfo {author} {\bibfnamefont {Di}~\bibnamefont {Xiao}},\ }\bibfield
  {title} {\enquote {\bibinfo {title} {Optical selection rule of excitons in
  gapped chiral fermion systems},}\ }\href {\doibase
  10.1103/PhysRevLett.120.077401} {\bibfield  {journal} {\bibinfo  {journal}
  {Phys. Rev. Lett.}\ }\textbf {\bibinfo {volume} {120}},\ \bibinfo {pages}
  {077401} (\bibinfo {year} {2018})}\BibitemShut {NoStop}%
\bibitem [{Note7()}]{Note7}%
  \BibitemOpen
  \bibinfo {note} {Non resonant SHG experiments in \protect \textit {h}-BN\ are
  described in [\protect \rev@citealpnum {Kim2013,Li2013}] whereas \protect
  \textit {ab initio}\ calculations are presented in [\protect \rev@citealpnum
  {Gruning2014}].}\BibitemShut {Stop}%
\bibitem [{\citenamefont {Wang}\ \emph {et~al.}(2015)\citenamefont {Wang},
  \citenamefont {Marie}, \citenamefont {Gerber}, \citenamefont {Amand},
  \citenamefont {Lagarde}, \citenamefont {Bouet}, \citenamefont {Vidal},
  \citenamefont {Balocchi},\ and\ \citenamefont {Urbaszek}}]{Wang2015}%
  \BibitemOpen
  \bibfield  {author} {\bibinfo {author} {\bibfnamefont {G.}~\bibnamefont
  {Wang}}, \bibinfo {author} {\bibfnamefont {X.}~\bibnamefont {Marie}},
  \bibinfo {author} {\bibfnamefont {I.}~\bibnamefont {Gerber}}, \bibinfo
  {author} {\bibfnamefont {T.}~\bibnamefont {Amand}}, \bibinfo {author}
  {\bibfnamefont {D.}~\bibnamefont {Lagarde}}, \bibinfo {author} {\bibfnamefont
  {L.}~\bibnamefont {Bouet}}, \bibinfo {author} {\bibfnamefont
  {M.}~\bibnamefont {Vidal}}, \bibinfo {author} {\bibfnamefont
  {A.}~\bibnamefont {Balocchi}}, \ and\ \bibinfo {author} {\bibfnamefont
  {B.}~\bibnamefont {Urbaszek}},\ }\bibfield  {title} {\enquote {\bibinfo
  {title} {Giant enhancement of the optical second-harmonic emission of
  ${\mathrm{wse}}_{2}$ monolayers by laser excitation at exciton resonances},}\
  }\href {\doibase 10.1103/PhysRevLett.114.097403} {\bibfield  {journal}
  {\bibinfo  {journal} {Phys. Rev. Lett.}\ }\textbf {\bibinfo {volume} {114}},\
  \bibinfo {pages} {097403} (\bibinfo {year} {2015})}\BibitemShut {NoStop}%
\bibitem [{\citenamefont {Ventura}\ \emph {et~al.}(2017)\citenamefont
  {Ventura}, \citenamefont {Passos}, \citenamefont {Lopes~dos Santos},
  \citenamefont {Viana Parente~Lopes},\ and\ \citenamefont
  {Peres}}]{Ventura2017}%
  \BibitemOpen
  \bibfield  {author} {\bibinfo {author} {\bibfnamefont {G.~B.}\ \bibnamefont
  {Ventura}}, \bibinfo {author} {\bibfnamefont {D.~J.}\ \bibnamefont {Passos}},
  \bibinfo {author} {\bibfnamefont {J.~M.~B.}\ \bibnamefont {Lopes~dos
  Santos}}, \bibinfo {author} {\bibfnamefont {J.~M.}\ \bibnamefont {Viana
  Parente~Lopes}}, \ and\ \bibinfo {author} {\bibfnamefont {N.~M.~R.}\
  \bibnamefont {Peres}},\ }\bibfield  {title} {\enquote {\bibinfo {title}
  {Gauge covariances and nonlinear optical responses},}\ }\href {\doibase
  10.1103/PhysRevB.96.035431} {\bibfield  {journal} {\bibinfo  {journal} {Phys.
  Rev. B}\ }\textbf {\bibinfo {volume} {96}},\ \bibinfo {pages} {035431}
  (\bibinfo {year} {2017})}\BibitemShut {NoStop}%
\bibitem [{\citenamefont {Kim}\ \emph {et~al.}(2013)\citenamefont {Kim},
  \citenamefont {Brown}, \citenamefont {Graham}, \citenamefont {Hovden},
  \citenamefont {Havener}, \citenamefont {McEuen}, \citenamefont {Muller},\
  and\ \citenamefont {Park}}]{Kim2013}%
  \BibitemOpen
  \bibfield  {author} {\bibinfo {author} {\bibfnamefont {Cheol-Joo}\
  \bibnamefont {Kim}}, \bibinfo {author} {\bibfnamefont {Lola}\ \bibnamefont
  {Brown}}, \bibinfo {author} {\bibfnamefont {Matt~W.}\ \bibnamefont {Graham}},
  \bibinfo {author} {\bibfnamefont {Robert}\ \bibnamefont {Hovden}}, \bibinfo
  {author} {\bibfnamefont {Robin~W.}\ \bibnamefont {Havener}}, \bibinfo
  {author} {\bibfnamefont {Paul~L.}\ \bibnamefont {McEuen}}, \bibinfo {author}
  {\bibfnamefont {David~A.}\ \bibnamefont {Muller}}, \ and\ \bibinfo {author}
  {\bibfnamefont {Jiwoong}\ \bibnamefont {Park}},\ }\bibfield  {title}
  {\enquote {\bibinfo {title} {Stacking order dependent second harmonic
  generation and topological defects in h-bn bilayers},}\ }\href {\doibase
  10.1021/nl403328s} {\bibfield  {journal} {\bibinfo  {journal} {Nano Letters}\
  }\textbf {\bibinfo {volume} {13}},\ \bibinfo {pages} {5660--5665} (\bibinfo
  {year} {2013})}\BibitemShut {NoStop}%
\bibitem [{\citenamefont {Li}\ \emph {et~al.}(2013)\citenamefont {Li},
  \citenamefont {Rao}, \citenamefont {Mak}, \citenamefont {You}, \citenamefont
  {Wang}, \citenamefont {Dean},\ and\ \citenamefont {Heinz}}]{Li2013}%
  \BibitemOpen
  \bibfield  {author} {\bibinfo {author} {\bibfnamefont {Yilei}\ \bibnamefont
  {Li}}, \bibinfo {author} {\bibfnamefont {Yi}~\bibnamefont {Rao}}, \bibinfo
  {author} {\bibfnamefont {Kin~Fai}\ \bibnamefont {Mak}}, \bibinfo {author}
  {\bibfnamefont {Yumeng}\ \bibnamefont {You}}, \bibinfo {author}
  {\bibfnamefont {Shuyuan}\ \bibnamefont {Wang}}, \bibinfo {author}
  {\bibfnamefont {Cory~R.}\ \bibnamefont {Dean}}, \ and\ \bibinfo {author}
  {\bibfnamefont {Tony~F.}\ \bibnamefont {Heinz}},\ }\bibfield  {title}
  {\enquote {\bibinfo {title} {Probing symmetry properties of few-layer mos2
  and h-bn by optical second-harmonic generation},}\ }\href {\doibase
  10.1021/nl401561r} {\bibfield  {journal} {\bibinfo  {journal} {Nano Letters}\
  }\textbf {\bibinfo {volume} {13}},\ \bibinfo {pages} {3329--3333} (\bibinfo
  {year} {2013})}\BibitemShut {NoStop}%
\bibitem [{\citenamefont {Gr\"uning}\ and\ \citenamefont
  {Attaccalite}(2014)}]{Gruning2014}%
  \BibitemOpen
  \bibfield  {author} {\bibinfo {author} {\bibfnamefont {M.}~\bibnamefont
  {Gr\"uning}}\ and\ \bibinfo {author} {\bibfnamefont {C.}~\bibnamefont
  {Attaccalite}},\ }\bibfield  {title} {\enquote {\bibinfo {title} {Second
  harmonic generation in $h$-bn and mos${}_{2}$ monolayers: Role of
  electron-hole interaction},}\ }\href {\doibase 10.1103/PhysRevB.89.081102}
  {\bibfield  {journal} {\bibinfo  {journal} {Phys. Rev. B}\ }\textbf {\bibinfo
  {volume} {89}},\ \bibinfo {pages} {081102} (\bibinfo {year}
  {2014})}\BibitemShut {NoStop}%
\end{thebibliography}%

\end{document}